\documentclass[a4paper,fleqn,usenatbib]{mnras}

\usepackage{mathptmx}
\usepackage[T1]{fontenc}
\usepackage{ae,aecompl}

\usepackage{graphicx}	% Including figure files
\usepackage{amsmath}	% Advanced maths commands
\usepackage{amssymb}	% Extra maths symbols
\usepackage{pdflscape}

\title[Outflow energetics with variable BALs and \ion{P}{V}]{Quasar outflow energetics from broad absorption line variability}

\author[S. M. McGraw et al.]{S. M. McGraw,$^{1,2}$\thanks{E-mail: mcgrawsean01@gmail.com (SMM)}
J. C. Shields,$^{2}$\thanks{E-mail: shieldj1@ohio.edu (JCS)}
F. W. Hamann,$^{3}$
D. M. Capellupo,$^{4}$
\newauthor
and H. Herbst$^{5}$
\\
$^{1}$Department of Astronomy \& Astrophysics, The Pennsylvania State University, University Park, PA 16802, USA \\
$^{2}$Department of Physics and Astronomy, Ohio University, Athens, OH 45701, USA \\
$^{3}$Department of Physics and Astronomy, University of California Riverside, Riverside, CA 92507, USA \\
$^{4}$Department of Physics, McGill University, Montreal, Quebec H3A 2T8, Canada\\
$^{5}$Department of Astronomy, University of Florida, Gainesville, FL 32611, USA
}

\date{Accepted XXX. Received YYY; in original form ZZZ}

\pubyear{2017}

\begin{document}
\label{firstpage}
\pagerange{\pageref{firstpage}--\pageref{lastpage}}
\maketitle

\begin{abstract}

Quasar outflows have long been recognized as potential contributors to the co-evolution between supermassive black holes (SMBHs) and their host galaxies. The role of outflows in AGN feedback processes can be better understood by placing observational constraints on wind locations and kinetic energies. We utilize broad absorption line (BAL) variability to investigate the properties of a sample of 71 BAL quasars with \ion{P}{V} broad absorption. The presence of \ion{P}{V} BALs indicates that other BALs like \ion{C}{IV} are saturated, such that variability in those lines favours clouds crossing the line of sight. We use these constraints with measurements of BAL variability to estimate outflow locations and energetics. Our data set consists of multiple-epoch spectra from the Sloan Digital Sky Survey and MDM Observatory. We detect significant (4$\sigma$) BAL variations from 10 quasars in our sample over rest frame time-scales between $\le0.2-3.8$\,yr. Our derived distances for the 10 variable outflows are nominally $\la1-10$\,pc from the SMBH using the transverse-motion scenario, and $\la100-1000$\,pc from the central source using ionization-change considerations. These distances, in combination with the estimated high outflow column densities (i.e. $N_{\rmn{H}}\ga10^{22}$\,cm$^{-2}$), yield outflow kinetic luminosities between $\sim0.001-1$ times the bolometric luminosity of the quasar, indicating that many absorber energies within our sample are viable for AGN feedback.

\end{abstract}

\begin{keywords}
quasars: absorption lines -- quasars: general -- galaxies: active
\end{keywords}

\section{Introduction}

Quasar outflows are important candidates for active galactic nuclei (AGN) feedback processes that contribute to the co-evolution between supermassive black holes (SMBHs) and their host galaxies (see e.g., \citealt{kor13}). High column density outflows sufficiently far from the SMBH can inject energy and momentum into the ISM, regulating star formation and accretion on to the SMBH (see e.g., \citealt{fab12}). AGN feedback may also be responsible for the observed correlations between the SMBH mass and properties of the surrounding galaxy bulge (e.g. the $M$--$\sigma$ relation, \citealt{fer00}; \citealt{geb00}).

Understanding the role of outflows in AGN feedback requires estimates of mass flow rate, kinetic energy luminosity, and momentum flux, which depend on the outflow's speed, column density, and distance from the SMBH. Outflow column densities and distances can be extracted from rest-frame UV quasar spectra. Blue-shifted broad absorption lines (BALs) are indicative of quasar outflows and trace gas with a range of ionization states.

BAL variability studies can be used to constrain the distance of the outflow from the SMBH, thereby enabling estimates of feedback quantities given knowledge of the column density and global covering fraction of the absorber. Distances derived using this approach depend on the interpretation of the absorption line variability. One scenario involves absorbers crossing the line of sight (LOS) due to transverse motions and thereby producing a change in coverage of the background light source. In this case the geometry of the absorber and background light source along with the measured variability time yield a crossing speed, which can then constrain the distance from the SMBH if the crossing speeds are roughly Keplerian \cite[see Section 4.2.1 and][for more discussion]{mor17}. A second scenario consists of the outflowing gas undergoing an ionization change. The BAL variability time-scale in this situation is dictated by the time-scale of fluctuations in the ionizing radiation field and by the recombination time of the gas, which allows for limits on the outflow density and distance through knowledge of the ionization state of the absorber. Other scenarios are possible but potentially problematic, as discussed by \citet{cap13}.

There are conflicting predictions for the locations of BAL outflows relative to the SMBH. \citet{mur95} and \citet{dek95} proposed theoretical models that associate the BAL gas with winds accelerated off the accretion disc at sub-parsec scales from the SMBH (see also \citealt{pro00}; \citealt{pro04}; \citealt{fuk10}). In contrast \citet{fau12a} and \citet{fau12b} suggested that winds from the AGN shock and accelerate ISM gas that travels out to kilo-parsec distances from the SMBH, creating the observed FeLoBALs at such scales. 

BAL variability studies have generally constrained outflowing gas to be within hundreds of parsecs of the SMBH. Work involving \ion{C}{IV} and \ion{Si}{IV} BAL variability has estimated outflow distances within 0.1\,pc \citep{moe09} and 10\,pc \citep{cap11,cap12,cap13} of the SMBH by attributing the changes to absorbers crossing the LOS. \citet{fil13} found that 50--60 per cent of both \ion{C}{IV} and \ion{Si}{IV} BALs in their sample of 291 BAL quasars exhibited variability over multi-year time-scales, and concluded that such a high fraction is consistent with models where most BAL winds arise between $\sim$0.01--1\,pc from the SMBH. \citet{bar94} detected variability in \ion{C}{IV}, \ion{Si}{IV}, \ion{N}{V}, and \ion{Al}{III} BALs and constrained the outflows to be within a few hundred parsecs of the SMBH using an ionization change interpretation. \ion{Mg}{II} and \ion{Fe}{II} BAL variability studies have constrained the absorbers to be within parsecs and tens of parsecs of the SMBH based on models involving outflows crossing the LOS (\citealt{hal11}; \citealt{mcg15}) and photoionization models \citep{zha15}. \citet{viv12}, however, concluded that variable \ion{Fe}{III} and \ion{Al}{III} BALs in their sample reside in the ISM of the host galaxy using ionization considerations (see also \citealt{zha15}). Outflow kinetic luminosities derived from variability studies span a wide range, from $\sim$0.01 to 2 per cent of the bolometric luminosity of the quasar $L_{\rmn{bol}}$, depending on the available data (e.g., \citealt{cap14}; \citealt{gri15}).

Observational results on outflow energetics have also utilized photoionization models in conjunction with density-sensitive absorption lines to estimate the absorber's column density and distance from the SMBH. Examples of ions used to estimate the absorbing gas density include \ion{S}{IV} (\citealt{bor13}; \citealt{cha15a}), \ion{He}{I}$^*$ with \ion{Fe}{II} \citep{luc14}, \ion{O}{IV} \citep{ara13}, and \ion{Si}{II} (\citealt{moe09}; \citealt{dun10}). Outflow distances derived using this approach range from tens of parsecs to kilo-parsecs from the SMBH. These studies estimate outflow kinetic luminosities between $\sim$0.1$-$5 per cent of $L_{\rmn{bol}}$, which are broadly consistent with models that propose thresholds of 0.5$-$5 per cent of $L_{\rmn{bol}}$ for outflows to contribute to feedback (\citealt{sca04}; \citealt{hop10}). 

The \ion{P}{V} $\lambda\lambda$1117,1128 BAL is of particular interest in understanding the contribution of quasar outflows for AGN feedback since it traces high column density absorbers. Phosphorus is $\sim$1000 times less abundant in the Sun than carbon \citep{asp09}; if quasar outflows exhibiting \ion{P}{V} absorption have roughly solar relative abundances, then the \ion{C}{IV} BAL optical depth is between $\sim$100--1000 times greater than the \ion{P}{V} BAL optical depth (\citealt{ham98}; \citealt{lei11}; \citealt{bor12}). Measured BAL optical depths do not represent the true optical depths in high column density outflows of solar abundances because the absorber partially covers the continuum source (see e.g., \citealt{ham99}; \citealt{ara99,ara01}).

Few studies exist that constrain outflow distances using BAL variability and place limits on column densities using a sensitive diagnostic such as the \ion{P}{V} BAL; demographics of outflow energies and momenta are therefore likely incomplete. Capellupo et al. (2014) used the \ion{P}{V} BAL detected in Q1413+1143 to constrain the outflow hydrogen column density in units of cm$^{-2}$ to be log $N_{\rmn{H}} \ga 22.3$ assuming solar abundances (depending on the specific line strengths and widths, see also \citealt{mor17}). Detection of coordinated variability from both the \ion{P}{V} BAL and a saturated \ion{C}{IV} BAL that partially covers the continuum source provides strong evidence for interpretations involving absorbers crossing the LOS and allowed Capellupo et al. (2014) to place an upper limit of $\sim$3.5\,pc for the outflow distance from the SMBH. Capellupo et al. (2014) then used the column density and distance constraints to estimate the kinetic luminosity of the outflow to be $\sim$2 per cent of $L_{\rmn{bol}}$, concluding that the absorber might be relevant to feedback with the host galaxy.

We have assembled multiple-epoch spectra of 71 BAL quasars with \ion{P}{V} broad absorption for the purpose of estimating quasar outflow energetics. This sample expands on the work of Capellupo et al. (2014) by providing a significant increase in the number of sources analysed, and probes BAL variability over a range of time-scales using quantitative criteria in order to understand the viability of quasar outflows for feedback.

\section{Data}

Our sample of 71 BAL quasars consists of 68 objects from \citet{cap17} and 3 additional sources with similar attributes (122654--005430, 220359+005901, 221326+003846). \citet{cap17} detected 81 `definite' and very conservatively found 86 `probable' cases of \ion{P}{V} $\lambda\lambda$1117,1128 broad absorption by visual inspection of 2694 BAL quasars from the Sloan Digital Sky Survey (SDSS) data release 9. To distinguish from Ly$\alpha$ forest absorption lines at similar wavelengths, the selection process identified strong \ion{P}{V} BALs that exhibit similar velocity shifts and profiles as the \ion{C}{IV} $\lambda\lambda$1548,1550 and \ion{Si}{IV} $\lambda\lambda$1393,1402 BALs. Due to these selection criteria, our sample favours strong BALs with strong \ion{P}{V} absorption. The 68 objects from \citet{cap17} used for this analysis consists of 32 `definite' and 36 `probable' \ion{P}{V} BAL quasars, and were selected based on availability of multiple-epoch spectra taken from SDSS Data Release 7 (DR7) and 12 (DR12) and from the 2.4-m Hiltner telescope at MDM Observatory.

Observations at MDM Observatory were carried out for 14 of the 71 BAL quasars using the Boller and Chivens CCD Spectrograph (CCDS).\footnote{http://www.astronomy.ohio-state.edu/MDM/CCDS/} Integrated exposure times during a single run ranged from 2.5 to 7\,h for a given source at an average airmass of 1.2. We utilized a slit width of 1\,arcsec during all observations and rotated the slit to the parallactic angle to minimize effects of atmospheric dispersion. Spectra were processed using the Image Reduction and Analysis Facility (\textsc{iraf}, see \citealt{mcg15} for details).

Table~1 lists relevant information for the 71 BAL quasars in our sample including redshift, absolute $i$-band magnitude, radio loudness, SMBH mass, bolometric luminosity, Eddington ratio, dates of observation, wavelength coverage, and BAL variability. Our sample consists of quasars with redshifts between 2.4 and 4.4 and absolute $i$-band magnitudes between --29.1 and --25.8. There are 8 quasars in our sample that were detected in the Faint Images of the Radio Sky at Twenty-Centimeters (FIRST) survey, and yield radio loudness parameters  (i.e. $R \equiv f_{6\rmn{cm}}/f_{\rmn{2500\,A}}$) between 1 and 747. Radio loudness parameters were taken from \citet{she11} when available or were estimated using the FIRST integrated flux densities at 20\,cm along with measured rest-frame flux densities at 2500\,\AA. We follow \citet{she11} to determine the rest-frame flux density at 6\,cm by using the flux density at 20\,cm and assuming a power law with index $\alpha_{\nu}=-0.5$. 

\begin{table*}
	\centering
	\caption{BAL quasar sample}
	\label{tab:example_table}
	\begin{tabular}{cccccccccccc}
		\hline
		Name & $z^a$ & $M_{i}^b$ & $R^c$ & log $M_{\bullet}^d$ & log $L_{\rmn{bol}}^e$ & log $\frac{L_{\rmn{bol}}}{L_{\rmn{Edd}}}^f$ & DR7$^g$ 		& DR12 & MDM & Coverage$^h$ & Variability$^i$ \\
		\hline
		J001610.79+013608.0 & 2.83 & --27.0 & -- & -- & 46.4 & -- & -- & 55511 & -- & 930--2700 & N \\
		&&&&&&&& 55528 &&& \\
		J001824.95+001525.8 & 2.43 & --26.7 & -- & 8.4 & 46.6 & 0.0 & 51816 & 55480 & -- & 1110--2690 & T \\
		&&&&&&& 51900 &&&& \\
		J002417.61+000846.3 & 3.96 & --27.3 & -- & -- & 46.9 & -- & 52203 & 55447 & -- & 720--2090 & N \\
		&&&&&&&& 55480 &&& \\
		J003859.34--004252.3 & 2.50 & --26.1 & -- & -- & 46.2 & -- & 52261 & 55182 & -- & 1020--2950 & T \\
		&&&&&&&& 55184 &&& \\
		&&&&&&&& 55186 &&& \\
		&&&&&&&& 55444 &&& \\
		J010338.43--020047.2 & 2.84 & --27.4 & -- & -- & 46.9 & -- & -- & 55541 & -- & 940--2710 & T \\
		&&&&&&&& 55811 &&& \\
		J013652.52+122501.5 & 2.40 & --28.5 & 8 & 8.9 & 47.2 & 0.2 & -- & 55614 & 56304 & 1050--1850 & Y \\
		J013802.07+012424.4 & 2.53 & --28.4 & -- & -- & 47.2 & -- & -- & 55501 & 56306 & 990--1730 & T \\
		J014025.63+002707.9 & 2.54 & --26.9 & -- & 9.0 & 46.5 & --0.6 & -- & 55205 & -- & 1020--2930 & T \\
		&&&&&&&& 55483 &&& \\
		J014141.32+011205.7 & 3.14 & --26.3 & -- & 8.7 & 46.3 & --0.5 & -- & 55205 & -- & 870--2500 & N \\
		&&&&&&&& 55444 &&& \\
		J015032.87+143425.6 & 4.18 & --27.5 & -- & -- & 47.0 & -- & 51820 & 55591 & -- & 740--1770 & N \\
		J022122.51--044658.8 & 2.64 & --26.7 & -- & 8.6 & 46.4 & --0.2 & -- & 55534 & -- & 980--2840 & T \\
		&&&&&&&& 56217 &&& \\
		J025000.59--002431.0 & 2.78 & --26.7 & -- & 8.6 & 46.6 & --0.1 & -- & 55450 & -- & 940--2740 & T \\
		&&&&&&&& 55476 &&& \\
		J025042.45+003536.7 & 2.39 & --27.9 & -- & 8.9 & 47.0 & 0.0 & 51816 & 55450 & 56563 & 1060--3060 & Y \\
		&&&&&&& 51877 & 55476 &&& \\
		&&&&&&& 52177 &&&& \\
		&&&&&&& 52965 &&&& \\
		&&&&&&& 52973 &&&& \\
		J072444.07+392711.8 & 2.46 & --26.9 & 12 & -- & 46.8 & -- & -- & 55240 & 56306 & 1030--1810 & T \\
		J073656.27+440308.7 & 2.70 & --26.7 & -- & -- & 46.6 & -- & -- & 55481 & -- & 970--2800 & N \\
		&&&&&&&& 56326 &&& \\
		J073751.52+455140.6 & 2.40 & --26.9 & -- & 9.1 & 46.5 & --0.7 & -- & 55176 & -- & 1060--3060 & T \\
		&&&&&&&& 56218 &&& \\
		&&&&&&&& 56328 &&& \\
		J075014.40+432635.2 & 3.20 & --27.8 & -- & -- & 47.0 & -- & -- & 55182 & -- & 850--2460 & T \\
		&&&&&&&& 55481 &&& \\
		&&&&&&&& 56323 &&& \\
		J080029.38+124836.9 & 3.08 & --27.4 & -- & 9.1 & 46.5 & --0.7 & 53674 & 55603 & -- & 930--2260 & T \\
		J080040.95+160913.0 & 3.57 & --28.1 & 7 & 9.4 & 47.1 & --0.4 & 53350 & 55569 & -- & 840--2020 & N \\
		J081003.92+522507.7 & 3.88 & --28.4 & -- & 9.7 & 47.4 & --0.4 & -- & 55542 & -- & 740--2130 & N \\
		&&&&&&&& 55559 &&& \\
		J081208.61+534800.5 & 2.60 & --27.3 & -- & -- & 46.9 & -- & -- & 55540 & -- & 990--2860 & T \\
		&&&&&&&& 55563 &&& \\
		J081410.13+323225.1 & 3.62 & --27.0 & -- & 9.1 & 46.7 & --0.5 & 52318 & 55506 & -- & 820--1990 & T \\
		&&&&&&& 54578 &&&& \\
		&&&&&&& 54774 &&&& \\
		J081608.28+210213.2 & 3.03 & --28.0 & -- & 9.2 & 47.1 & --0.2 & 53330 & 55592 & -- & 950--2290 & T \\
		J082227.60+404153.8 & 2.95 & --26.2 & -- & 8.9 & 46.4 & --0.6 & -- & 55272 & -- & 910--2610 & N \\
		&&&&&&&& 55505 &&& \\
		J082543.23+383829.2 & 3.01 & --26.7 & -- & 8.1 & 46.6 & 0.4 & 52615 & 55272 & -- & 950--2300 & N \\
		J090035.30+040846.4 & 2.86 & --26.3 & -- & 9.7 & 46.6 & --1.2 & -- & 55277 & -- & 930--2680 & T \\
		&&&&&&&& 55535 &&& \\
		J094431.33+033411.6 & 3.00 & --27.3 & -- & 8.8 & 46.6 & --0.4 & 52266 & 55631 & -- & 890--2580 & T \\
		&&&&&&&& 55672 &&& \\
		\hline
		\multicolumn{12}{l}{$^a$ Emission-line redshifts taken from \citet{hew10}, if available, or from \citet{par14}}\\
		\multicolumn{12}{l}{$^b$ Absolute $i$-band magnitudes corrected to $z=2$ \citep{she11}}\\
		\multicolumn{12}{l}{$^c$ Radio loudness parameters (i.e. $R \equiv \frac{f_{6\rmn{cm}}}{f_{\rmn{2500\,A}}}$, see Section 2 for details)}\\
		\multicolumn{12}{l}{$^d$ Virial SMBH mass estimates in solar units (see Section 2 for details)}\\
		\multicolumn{12}{l}{$^e$ Bolometric luminosities in erg\,s$^{-1}$ (see Section 2 for details)}		\\
		\multicolumn{12}{l}{$^f$ Eddington ratios taken from \citet{she11}, if available, or calculated using our estimates of $M_{\bullet}$ and $L_			{\rmn{bol}}$}\\
		\multicolumn{12}{l}{$^g$ Spectroscopic observations listed as Modified Julian Dates (MJDs)}\\
		\multicolumn{12}{l}{$^h$ Quasar-frame wavelength intervals in \AA \ with at least two spectroscopic observations to probe BAL variability}\\
		\multicolumn{12}{l}{$^i$ Significance of BAL variability categorized as yes (Y), tentative (T), or no (N); see Section 3.1 for details}\\
	\end{tabular}
\end{table*}

\begin{table*}
	\centering
	\contcaption{}
	\label{tab:example_table}
	\begin{tabular}{cccccccccccc}
		\hline
		Name & $z$ & $M_{i}$ & $R$ & log $M_{\bullet}$ & log $L_{\rmn{bol}}$ & log $\frac{L_{\rmn{bol}}}{L_{\rmn{Edd}}}$ & DR7 & DR12 & MDM & Coverage & Variability \\
		\hline
		J094633.97+365516.8 & 2.84 & --27.5 & -- & -- & 46.9 & -- & 53035 & 55590 & -- & 990--2400 & N \\
		J095333.70+033623.7 & 3.29 & --27.2 & -- & -- & 46.3 & -- & 52286 & 55656 & -- & 880--2100 & T \\
		J095442.89+432512.0 & 2.50 & --27.1 & -- & 9.6 & 46.1 & --1.6 & 52703 & 55649 & -- & 1090--2630 & T \\
		&&&&&&& 52709 &&&& \\
		J101225.00+405753.3 & 3.17 & --27.0 & -- & 8.2 & 46.5 & 0.2 & 53034 & 55570 & -- & 910--2210 & T \\
		J101324.20+064900.3 & 2.77 & --27.6 & -- & 8.7 & 46.9 & 0.1 & 52641 & 55677 & 56304 & 950--2750 & T \\
		&&&&&&&& 55679 &&& \\
		J101412.56+394135.7 & 3.27 & --27.2 & -- & 9.3 & 46.3 & --1.1 & 53034 & 55570 & -- & 890--2150 & T \\
		J102154.01+051646.4 & 3.46 & --28.8 & -- & 10.5 & 47.3 & --1.3 & 52319 & 55652 & -- & 850--2070 & Y \\
		J102251.29+031529.3 & 3.59 & --26.6 & -- & 8.6 & 46.2 & --0.5 & 51999 & 55652 & -- & 830--2010 & N \\
		J102744.88+041737.5 & 2.68 & --25.9 & -- & -- & 46.6 & -- & -- & 55652 & -- & 970--2800 & N \\
		&&&&&&&& 55654 &&& \\
		J103242.71+433605.2 & 3.49 & --27.8 & -- & 9.4 & 47.0 & --0.5 & 52992 & 55651 & -- & 850--2050 & T \\
		J103958.20+061119.7 & 3.14 & --26.6 & 23 & 7.9 & 46.3 & 0.3 & 52643 & 55689 & -- & 920--2220 & N \\
		J104059.79+055524.3 & 2.45 & --27.0 & 747 & 9.3 & 46.6 & --0.8 & 52643 & 55648 & -- & 1100--2670 & Y \\
		J104247.56+061521.4 & 2.51 & --25.8 & -- & -- & 46.4 & -- & -- & 55689 & -- & 1010--2940 & Y \\
		&&&&&&&& 55928 &&& \\
		J104846.63+440710.8 & 4.41 & --28.1 & -- & 9.6 & 47.1 & --0.6 & 53050 & 55653 & -- & 700--1700 & T \\
		J105928.53+011417.2 & 2.63 & --27.4 & -- & 9.2 & 47.0 & --0.3 & -- & 55302 & -- & 990--2840 & T \\
		&&&&&&&& 55648 &&& \\
		J114056.81--002329.9 & 3.61 & --27.5 & -- & 8.9 & 47.0 & 0.0 & 51584 & 55572 & -- & 770--2240 & T \\
		&&&&&&& 51660 & 56016 &&& \\
		&&&&&&& 51959 &&&& \\
		J114548.38+393746.6 & 3.12 & --29.1 & -- & 9.4 & 47.6 & 0.1 & 53442 & 55659 & 56304 & 870--2230 & Y \\
		J114847.07+395544.8 & 2.99 & --27.2 & -- & 8.9 & 46.6 & --0.4 & 53386 & 55659 & -- & 950--2300 & T \\
		J120447.15+330938.7 & 3.64 & --28.7 & 1 & 9.5 & 46.9 & --0.7 & 53498 & 55603 & -- & 820--1980 & T \\
		J120704.75+033243.9 & 2.72 & --26.2 & -- & 8.6 & 46.4 & --0.3 & -- & 55631 & -- & 970--2780 & T \\
		&&&&&&&& 55652 &&& \\
		J120834.84+002047.7 & 2.70 & --28.3 & -- & 9.7 & 47.0 & --0.8 & 51616 & 55321 & -- & 1030--2490 & T \\
		&&&&&&& 51666 &&&& \\
		&&&&&&& 51999 &&&& \\
		&&&&&&& 54992 &&&& \\
		J121858.14+005053.7 & 3.08 & --26.8 & -- & 9.3 & 46.8 & --0.6 & -- & 55327 & -- & 880--2530 & T \\
		&&&&&&&& 55630 &&& \\
		J122654.39--005430.6 & 2.61 & --28.3 & -- & 9.4 & 47.2 & --0.2 & 51990 & 55212 & 56307 & 990--2550 & Y \\
		&&&&&&&&& 57037 && \\
		J131333.01--005114.3 & 2.95 & --27.5 & -- & 9.0 & 46.7 & --0.4 & 51585 & 55321 & -- & 960--2330 & T \\
		&&&&&&& 51985 &&&& \\
		J132004.70+363830.1 & 2.75 & --27.3 & -- & 9.4 & 46.5 & --1.0 & 53851 & 55597 & -- & 1020--2460 & T \\
		J132139.86--004151.9 & 3.12 & --28.1 & 42 & 8.5 & 47.2 & 0.6 & 51578 & 55599 & -- & 920--2230 & T \\
		&&&&&&& 51665 &&&& \\
		&&&&&&& 51984 &&&& \\
		J135536.89+320323.7 & 2.76 & --26.4 & -- & 8.6 & 46.4 & --0.2 & 54115 & 55274 & -- & 1010--2440 & T \\
		J140532.90+022957.3 & 2.83 & --28.4 & -- & 9.5 & 47.2 & --0.4 & 51993 & 55332 & 56426 & 930--2410 & Y \\
		&&&&&&&&& 57038 && \\
		J141906.28+362801.9 & 2.40 & --27.6 & 8 & -- & 46.7 & -- & -- & 55246 & 56426 & 1020--1780 & T \\
		J141934.63+050327.1 & 2.49 & --26.9 & -- & 8.5 & 46.7 & 0.0 & 52055 & 55654 & -- & 1100--2640 & T \\
		J143223.10--000116.4 & 2.48 & --26.3 & -- & 8.0 & 46.3 & 0.2 & 51637 & 55350 & -- & 1090--2640 & T \\
		&&&&&&& 51690 &&&& \\
		J143632.04+053958.9 & 2.98 & --26.8 & -- & 8.9 & 46.8 & --0.2 & -- & 55682 & -- & 900--2590 & N \\
		&&&&&&&& 55684 &&& \\
		J151102.01+012659.0 & 2.92 & --27.7 & -- & 9.3 & 46.8 & --0.6 & 51996 & 55629 & -- & 970--2350 & N \\
		J152316.17+335955.5 & 3.64 & --27.2 & -- & 9.0 & 46.8 & --0.3 & 52814 & 55738 & -- & 820--1980 & T \\
		J153252.96+023217.3 & 2.71 & --27.3 & -- & 8.6 & 46.4 & --0.2 & 52026 & 55358 & -- & 1030--2480 & N \\
		J154435.61--001928.0 & 2.66 & --27.6 & -- & 8.3 & 46.3 & 0.0 & 51691 & 55356 & -- & 1040--2510 & T \\
		J155514.85+100351.3 & 3.51 & --28.9 & -- & 10.1 & 47.2 & --1.0 & 54572 & 55706 & 56428 & 850--2030 & T \\
		J162554.62+322626.5 & 3.79 & --27.3 & -- & 8.5 & 46.8 & 0.2 & 53239 & 55750 & -- & 790--1920 & T \\
		J165053.78+250755.4 & 3.34 & --28.5 & -- & 9.2 & 47.3 & 0.0 & 54621 & 55685 & -- & 880--2120 & T \\
		J165710.56+233700.2 & 2.80 & --26.7 & -- & -- & 46.7 & -- & 52912 & 55685 & -- & 1010--2420 & N \\
		J220359.73+005901.9 & 2.86 & --28.5 & -- & 9.5 & 47.4 & --0.2 & -- & 55481 & 56563 & 930--1630 & T \\
		J221326.95+003846.0 & 2.52 & --27.5 & -- & 9.3 & 46.8 & --0.6 & 52964 & 55499 & 56565 & 1020--2620 & Y \\
		J230721.91+011118.0 & 2.77 & --27.5 & -- & -- & 47.0 & -- & -- & 55476 & 56568 & 950--1670 & Y \\
		J231923.83+004127.6 & 2.54 & --27.5 & -- & -- & 47.1 & -- & -- & 55444 & 56564 & 1030--1800 & T \\
		\hline
	\end{tabular}
\end{table*}

The range of SMBH masses in solar units, bolometric luminosities in erg\,s$^{-1}$, and Eddington ratios are from log $M_{\bullet}=7.9$ to 10.5, log $L_{\rmn{bol}}=46.1$ to 47.6, and log ($L_{\rmn{bol}}/L_{\rmn{edd}})=$ --1.6 to 0.6.\footnote{We note that single-epoch virial black hole masses and bolometric luminosities estimated using bolometric corrections suffer from significant systematic uncertainties (see e.g., \citealt{ric11}; \citealt{she13}). Since we are only obtaining order-of-magnitude estimates of BAL outflow locations and energetics, high precision masses and luminosities are not important for this study.} We have acquired between 2 and 8 spectra for a given source and probe BAL variability over rest frame time-scales from 0.5\,d to 3.8\,yr and rest-frame wavelengths ranging from 700 to 3060\,\AA. The virial SMBH masses were taken from \citet{she11}, if available, or were estimated using luminosities measured at 1350\,\AA \ along with full width at half-maxima (FWHMs) of the \ion{C}{IV} broad emission lines (BELs) taken from \citet{par14}. Following \citet{she11}, luminosities at 1350\,\AA \ and flux densities at 2500\,\AA \ were determined in a cosmology with $H_{\rmn{o}}=70$\,km\,s\,$^{-1}$\,Mpc$^{-1}$, $\Omega_{\rmn{M}}=0.3$, and $\Omega_{\rmn{\Lambda}}=0.7$, and were corrected for Galactic extinction using the dust maps from \citet{sch98} and the curve from \citet{car89} with $R_{V}=3.1$. Bolometric luminosities were taken from \citet{she11} when available or were calculated using the luminosities at 1350\,\AA \ and a bolometric correction factor of 3.81 from \citet{she11}.

The spectroscopic data are supplemented with synthesized $V$-band photometry, measured using images taken by an unfiltered CCD, from the Catalina Sky Survey (CSS) data release 2 (\citealt{dra09}, see Section 4.1). One quasar in our sample, 122654--005430, was observed using the Advanced CCD Imaging Spectrometer (ACIS) onboard the $Chandra$ X-ray Observatory, and we utilize the available data from the $Chandra$ Source Catalog (\citealt{eva10}, see Appendix B).  

Quasars observed during SDSS DR12 exhibit a known systematic offset in flux that is dependent on the airmass of the observation \citep{har16}. We account for this error in all derived quantities that involve measurements of flux density including radio loudness, SMBH mass, bolometric luminosity, and Eddington ratio. 

\section{Analysis}

\subsection{Absorption line variability}

Our goal is to detect BAL variability by utilizing quantitative criteria that involve comparing flux differences with propagated errors. For a given comparison between two spectra, the following operations were executed. The higher resolution spectrum was smoothed by a Gaussian of appropriate width to match the resolution of the other spectrum, wavelength and vacuum corrections were applied to CCDS spectra, all spectra were resampled to $70$\,km\,s$^{-1}$, and the spectrum acquired later in time was scaled to coincide with the  earlier spectrum.  This method is similar to the procedure described in \citet{mcg15}. Table 2 lists wavelength coverages, pixel widths, and spectroscopic resolutions for each of the three instruments used in this study.

\begin{table}
	\centering
	\caption{Spectroscopic parameters}
	\label{tab:example_table}
	\begin{tabular}{lccc}
		\hline
		Instrument & $\lambda$ coverage & Pixel width & Resolution \\
		& (\AA) & (km\,s\,$^{-1}$) & (km\,s\,$^{-1}$) \\
		\hline
		DR7 & 3800--9200 & 70 & 150 \\
		DR12 & 3600--10\,400 & 70 & 140 \\
		CCDS & 3200--7600 & 80 & 210 \\
		\hline
	\end{tabular}\\
The pixel width for CCDS and FWHM resolutions for all instruments are average estimates. Also note that the wavelength coverage for CCDS represents the limiting wavelengths observed across our sample.
\end{table}

Scaling one spectrum to coincide with another required fitting either a linear or cubic function to the ratio between the two spectra using wavelength intervals that do not exhibit BAL variability. Fig.~1 shows two representative examples of the ratio-fit procedure. Non-variable wavelength intervals (see orange intervals in Fig.~1) were determined by fitting the entire wavelength coverage and applying a sigma-clipping algorithm to reject localized, variable wavelength regions. The rejection algorithm involved computing the root mean squared (RMS) deviation between ratio values and the fit across the entire ratio spectrum, rejecting data points that deviate more than 3 times the RMS error from the fit\footnote{In a small number of comparisons we inspected the ratio spectra directly and altered the 3$\sigma$ threshold as needed to achieve a more reliable ratio fit.}, and repeating these steps until the fit converged to a solution (see solid red lines in Fig.~1). This procedure does not rely on pre-defined wavelength intervals to constrain the fit, and is therefore useful for our dataset that contains differing wavelength coverages between comparisons.

\begin{figure*}
	\includegraphics[width=1.8\columnwidth]{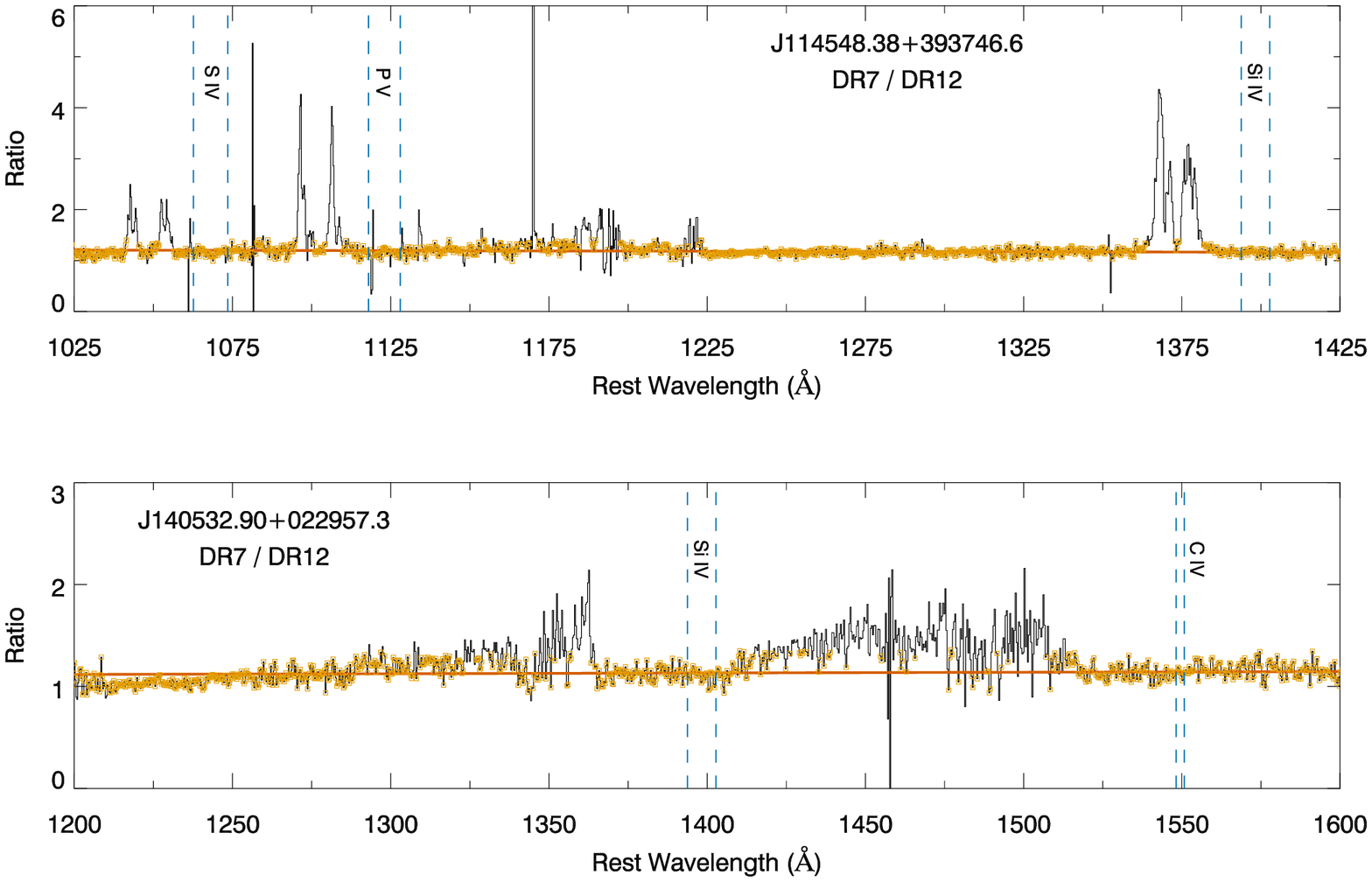}
	\caption{Representative examples of the ratio-fit procedure used for scaling spectra, which involves fitting the ratio between two spectra using non-variable wavelength intervals (see Section 3.1). Each panel lists the BAL quasar name and surveys (DR7 or DR12) associated with the two spectra used for calculating the ratio spectrum (black). Orange data intervals show the wavelength intervals that remained after the sigma-clipping process; these intervals were used to constrain the ratio fit (shown as the red solid line). Features in the displayed ratio spectra that are not associated with orange intervals likely represent absorption line variability. Vertical, blue, dashed lines show the locations of the S\thinspace\textsc{iv}, P\thinspace\textsc{v}, Si\thinspace\textsc{iv}, and C\thinspace\textsc{iv} doublets in the rest frame of the quasar.}
\end{figure*}

After scaling, one spectrum was subtracted from the other to create a difference spectrum. In order to detect absorption-line variations related to BAL outflows and to avoid detection of spurious changes across individual, unresolved pixels, we binned each difference spectrum by 2000 and 500\,km\,s\,$^{-1}$. The 2000\,km\,s$^{-1}$ bin width is the minimum representative width of a BAL (\citealt{wey91}), and is utilized to search for BAL variability across wider regions. The 500\,km\,s$^{-1}$ bin width is the minimum width that traces high-velocity winds (i.e., the mini-BAL outflows; see e.g. \citealt{ham04}), and is used to search for BAL changes across narrower intervals. The cumulative sigma ($\sigma$) in a given \emph{binned} difference value was calculated by adding the propagated random and systematic errors in quadrature on a pixel-by-pixel basis (see Fig.~2). Random errors within each bin were propagated in quadrature to determine the random error for each binned difference value. Systematic errors within each bin were averaged as opposed to propagated in quadrature since these uncertainties are not independent and random from one another.

Systematic errors include uncertainties in matching resolutions, wavelength calibration errors in CCDS spectra, and errors in the ratio fits used for scaling. The standard deviation of the width of the Gaussian smoothing function used to match resolutions was estimated to be 68 per cent of the amount between the minimum and maximum resolutions for each instrument. Uncertainties in wavelength for a given CCDS spectrum were estimated using night sky emission lines.\footnote{We measured line center wavelengths of the \ion{Hg}{I} $\lambda$4358, \ion{O}{I} $\lambda$5577, or \ion{O}{I} $\lambda$6300 emission lines.} Errors in the scaling function were estimated by generating distributions of the residuals between non-rejected ratio values and the fit over localized wavelength intervals. The error in the fit at each wavelength was quantified by adding in quadrature the mean and standard deviation of the mean of the corresponding residual distribution. 

BAL variability was determined by comparing binned flux differences with propagated errors, and we categorized each case as N (no), T (tentative), or Y (yes). Binned, absorption-line differences <3$\sigma$, either positive or negative, using both the 2000 and 500\,km\,s$^{-1}$ bin widths are categorized as N. Differences between 3 and 4$\sigma$ contaminated by potential emission-line changes are also placed in the N category. Tentative (T) variability constitutes binned, absorption-line differences between 3 and 4$\sigma$ using either the 2000 or the 500\,km\,s$^{-1}$ bin widths.\footnote{For example, an absorption line variation that is between 3 and 4$\sigma$ using the 2000\,km\,s$^{-1}$ bin width is considered part of the T category even if the same variation is $<3\sigma$ using the 500\,km\,s$^{-1}$ bin width.} Absorption-line differences that are >4$\sigma$ for only 1--2 bins, are very close to the 4$\sigma$ threshold, are potentially contaminated by emission-line changes, and/or generally appear marginal/spurious by eye are also placed in the T category. Variable (Y) binned, absorption-line difference values are >4$\sigma$ over more than 2 consecutive bins using either the 500 or the 2000\,km\,s$^{-1}$ bin widths, and appear convincing by eye. If any BAL for a given quasar exhibits variable (Y) absorption-line differences over our measured time-scales, then the variability category for that quasar is Y (the same rule applies to T quasars; see last column of Table~1).

\subsection{\ion{P}{V} column densities}

We utilize the \ion{P}{V} BAL, a tracer of high column density gas, to constrain the hydrogen column density of the absorber (see Section 4.3). In order to estimate the \ion{P}{V} column density, the \ion{P}{V} BAL apparent optical depth profiles were estimated and then integrated over appropriate velocity intervals. The measured \ion{P}{V} column densities are lower limits due to potential effects of large optical depths masked by partial coverage of the background light source \cite[e.g.,][and references therein]{mor17}.

The \ion{P}{V} BAL apparent optical depths were calculated using the residual fluxes within the troughs [i.e., \mbox{$\tau=-$ln$(F_{v}/F_{c,v})$}], where $F_v$ and $F_{c,v}$ are the fluxes of the spectrum and continuum, respectively, at a given velocity $v$. The continuum fluxes were estimated by fitting a power law to either the DR7 or DR12 spectrum in each source and minimizing a $\chi^2$ statistic (see Fig.~2). The \ion{Si}{IV}, \ion{C}{IV}, and \ion{Al}{III} BELs were fitted using Gaussians simultaneously with the power law function. The fits included wavelength intervals that were free of BALs and ranged from approximately 1260--1350, 1370--1480, 1530--1800, and 1870--2500\,\AA.

We integrated the \ion{P}{V} BAL apparent optical depth profiles using the following equation, in Gaussian units: \\
\begin{equation}
    N=\frac{m_{e}c}{\pi e^2 f \lambda}\int \tau (v) dv
\end{equation} \\
where $f$ is the oscillator strength, $\lambda$ is the rest wavelength, and $\tau(v)$ is the apparent optical depth at a given velocity \citep{sav91}. Rest wavelengths and oscillator strength values were taken from \citet{mor03}, and we treat the \ion{P}{V} $\lambda\lambda$1117,1128 doublet as a single line at $\lambda=1121$\,\AA \ with an average oscillator strength of $f=0.705$. For sources that showed non-variable \ion{P}{V} BALs, we integrated $\tau(v)$ in accordance with equation (1). For sources that showed variable \ion{P}{V} BALs, we integrated the optical depth difference to generate lower limits for the change in \ion{P}{V} column density.  

\begin{table}
	\centering
	\caption{Variable BAL properties}
	\label{tab:example_table}
	\begin{tabular}{ccccc}
		\hline
		Quasar name & BALs & $\Delta t$ & $v_{\rmn{LOS}}$ & Direction \\
		&& (yr) & (km\,s\,$^{-1}$) & \\
		\hline
		013652+122501 & \ion{P}{V} & 0.6 & --6200 & -- \\
		025042+003536 & \ion{Si}{IV} & 2.6 & --9400 & -- \\
		& \ion{C}{IV} & 2.6 & --9400 & -- \\
		102154+051646 & \ion{C}{IV} & 2.0 & --23\,900 & + \\
		104059+055524 & \ion{Si}{IV} & 2.4 & --7700 & + \\
		104247+061521 & \ion{Si}{IV} & 0.2 & --6800 & -- \\
		114548+393746 & \ion{N}{III} & 1.5 & --5800 & + \\
		& \ion{O}{VI} & 1.5 & --11\,000 & + \\
		& \ion{S}{IV} & 1.5 & --5700 & + \\
		& \ion{P}{V} & 1.5 & --5800 & + \\
		& \ion{H}{I} & 1.5 & --7100 & + \\
		& \ion{Si}{IV} & 1.5 & --5600 & + \\
		& \ion{C}{IV} & 1.5 & --7500 & + \\
		&& 1.5 & --11\,100 & + \\
		122654--005430 & \ion{Si}{IV} & 2.4 & --3000 & + \\
		&& 0.8 & --2500 & + \\
		&& 0.5 & --2500 & -- \\
		& \ion{C}{IV} & 2.4 & --4300 & + \\
		&& 2.4 & --9000 & + \\
		&& 2.4 & --13\,700 & + \\
		&& 2.4 & --17\,600 & + \\
		&& 0.8 & --17\,600 & + \\
		140532+022947 & \ion{P}{V} & 2.4 & --7400 & + \\
		& \ion{N}{V} & 2.4 & --24\,000 & + \\
		&& 0.8 & --18\,700 & -- \\
		& \ion{Si}{IV} & 2.4 & --8900 & + \\
		&& 0.8 & --10\,200 & -- \\
		& \ion{C}{IV} & 2.4 & --23\,000 & + \\
		&& 0.8 & --19\,200 & -- \\
		221326+003846 & \ion{C}{IV} & 2.0 & --10\,000 & + \\
		&& 0.8 & --8100 & -- \\
		230721+011118 & \ion{C}{IV} & 0.8 & --6900 & + \\
		\hline
	\end{tabular}\\
	Information listed for each BAL quasar includes (from left to right): ions associated with variable BALs, measured time-scales of BAL variability (upper limits that are limited by temporal sampling), LOS velocities corresponding to maximum flux changes within the BALs, and whether the BAL depths increased (+) or decreased (--). Multiple, distinct time-scales listed for the same ion are in chronological order. Multiple velocities for the same time-scale correspond to kinematically separate absorption troughs.
\end{table}

\section{Results and Discussion}

\subsection{Absorption line variability}

We detect definite (category Y, see Section 3.1) variability from any BAL (i.e., from \ion{C}{IV}, \ion{Si}{IV}, or \ion{P}{V}) in 10 out of 71 BAL quasars in our sample (see Appendix A for notes on individual objects). Spectroscopic comparisons in our sample probe rest-frame time-scales between 0.5\,d and 3.8\,yr, and we measure BAL variability time-scales over \mbox{$\le$ 0.2--3.8}\,yr for our Y sources. Tentative (T) variations from any BAL are detected in 44 sources, and 17 objects exhibit no (N) BAL changes. Table 3 lists the associated ion, measured variability time-scale, LOS velocity of the variable region, and variability direction for each variable (Y) BAL in our sample. BALs not presented in Table~3 show no significant absorption line changes over the listed time-scales. For example, there are 4 cases of \ion{P}{V} and/or \ion{Si}{IV} BAL variations with no associated \ion{C}{IV} BAL changes in the same time-scale comparison (see 013652+122501, 104059+055524, 104247+061521, and 0.5\,yr for 122654--005430 in Table~3). We detect 4 cases of \ion{C}{IV} BAL variability with non-variable \ion{P}{V} absorption at similar velocities (see 2.4\,yr for 122654--005430, 0.8\,yr for 140532+022947, along with 2.0 and 0.8\,yr for 221326+003846 in Table~3). There are 5 cases of \ion{C}{IV} BAL changes with likely no associated \ion{P}{V} absorption at similar velocities (see 102154+051646, 114548+393746, 2.4 and 0.8\,yr for 122654--005430, and 2.4\,yr for 140532+022947 in Table~3). In 2 cases (025042+003536 and 230721+011118; see Table~3) 

\begin{landscape}	
	\begin{figure}
		\centering
		\includegraphics[width=0.85\columnwidth]{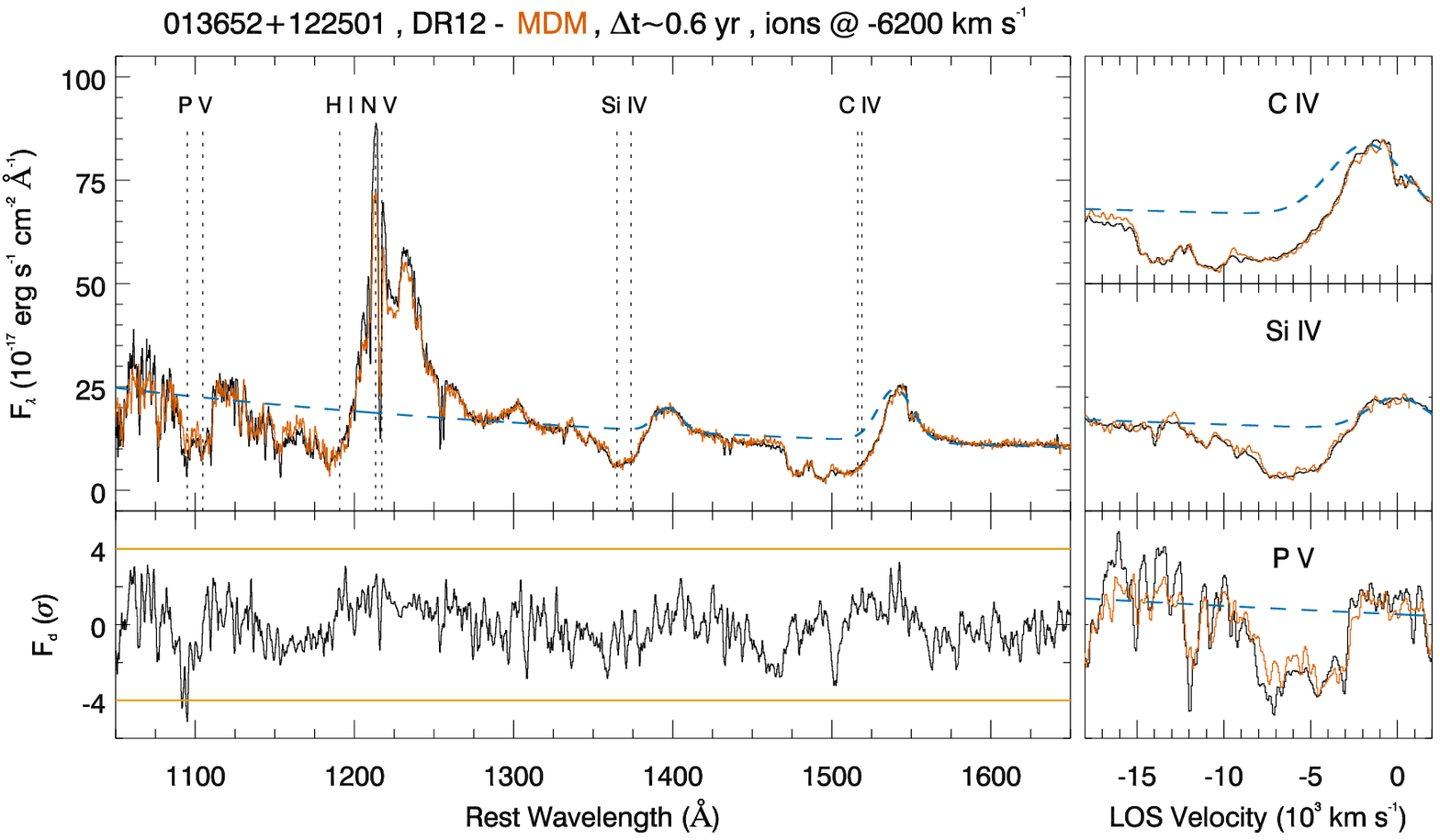}
	\end{figure}
\end{landscape} 

\clearpage

\begin{landscape}
	\begin{figure}
		\centering
		\includegraphics[width=0.85\columnwidth]{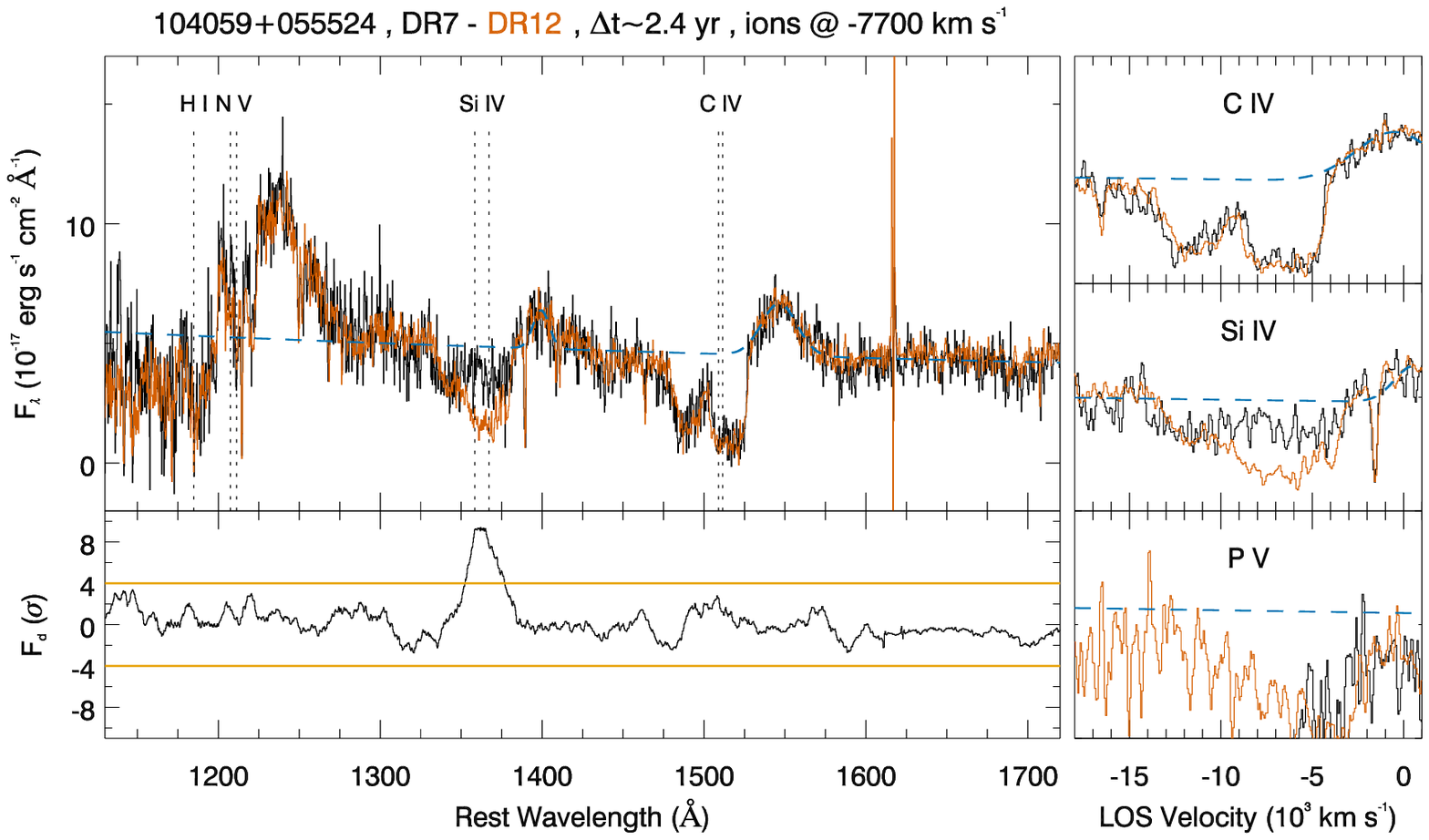}
	\end{figure}
\end{landscape} 

\clearpage

\begin{landscape}
	\begin{figure}
		\centering
		\includegraphics[width=0.85\columnwidth]{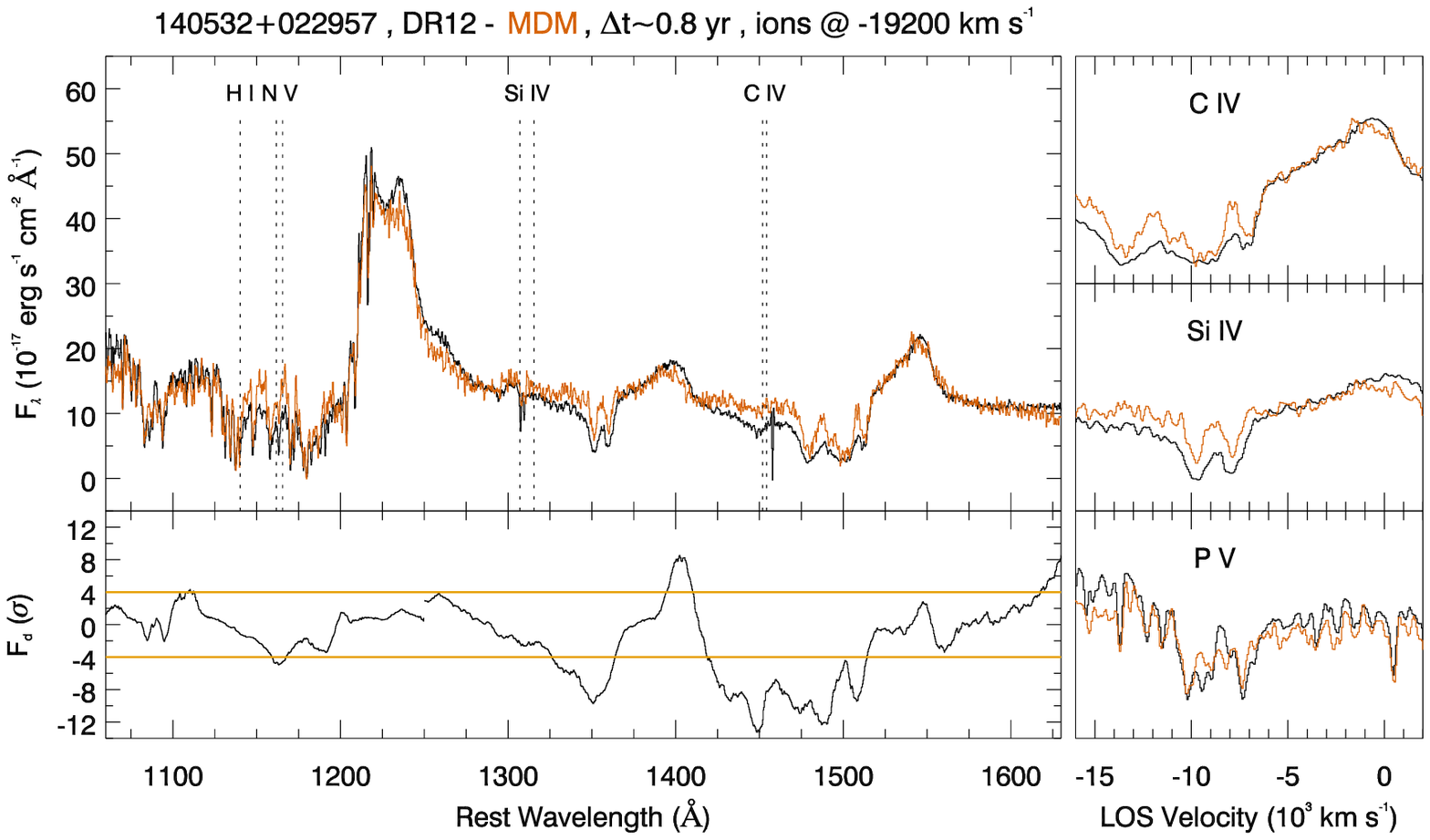}
	\end{figure}
\end{landscape} 

\clearpage

\setcounter{figure}{1}

\begin{landscape}
	\begin{figure}
		\centering
		\includegraphics[width=0.85\columnwidth]{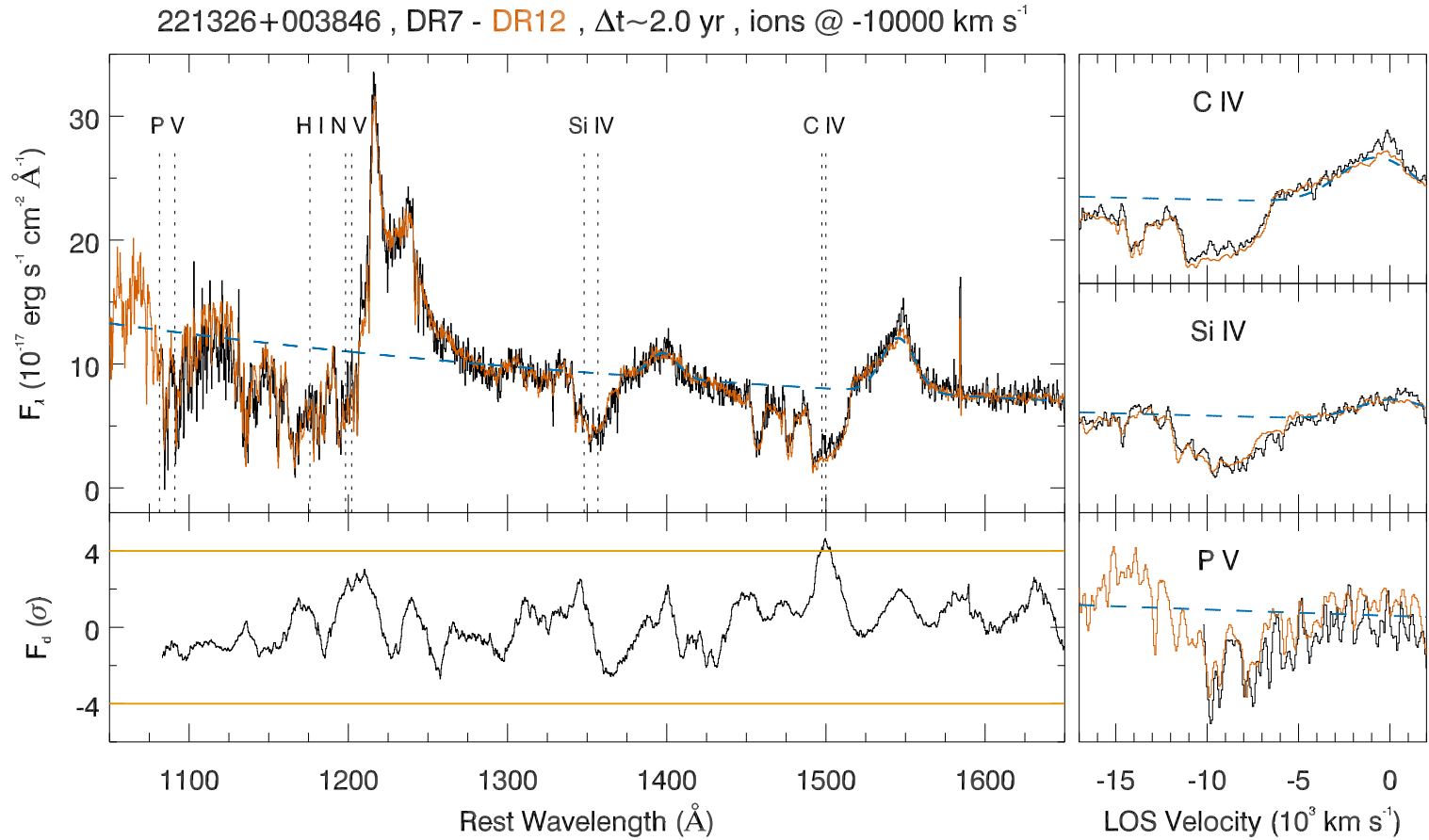}
		\caption{BAL quasar spectroscopic comparisons, revealing BAL variability from a range of ions. Details above the large panel include the BAL quasar name, instruments and colours used for the displayed spectra, rest-frame time-scale between observations, and the blue-shifted velocity of the labelled transitions (vertical, dotted lines). All vertical line labels are at the same velocity, and lie at a velocity that is centered on a variable (Y) interval within the BAL; these line labels do not correspond to the deepest portion of the BAL. The large panel displays two spectra over a large wavelength interval, and the three small panels on the right displays the two spectra in the vicinity of the \ion{C}{IV}, \ion{Si}{IV}, and \ion{P}{V} BALs on a common velocity and flux scale. Spectra in the small panels were smoothed using a 3-pixel wide boxcar for better presentation. The panel below the large panel displays the difference spectrum (in units of $\sigma$, see Section 3.1) for the two spectra, as well as the $\pm4\sigma$ thresholds (solid orange lines) used for detecting variable (Y) BALs. The order of the difference operation is the order of the instruments listed above the large panel. The dashed blue line in the large panel and small panels is the fit applied to either the DR7 or DR12 epoch and consists of a power law and Gaussians to model the continuum and BELs, respectively. All component figures are available in the online, supplementary material (see Fig.~2S).}
	\end{figure}
\end{landscape} 

\noindent it is unclear if the \ion{C}{IV} BAL variations are associated with \ion{P}{V} absorption at the same velocity.

The 10 Y sources exhibit BAL changes associated with resonance transitions from ions including \ion{N}{III} $\lambda\lambda$989,991, \ion{O}{VI} $\lambda\lambda$1031,1037, \ion{S}{IV} $\lambda\lambda$1062,1073, \ion{P}{V} $\lambda\lambda$1117,1128, \ion{H}{I} $\lambda$1215, \ion{N}{V} $\lambda\lambda$1238,1242, \ion{Si}{IV} $\lambda\lambda$1393,1402, and \ion{C}{IV} $\lambda\lambda$1548,1550. Tentative absorption line variability is detected from these transitions as well as from the \ion{C}{II} $\lambda\lambda$1334,1335 and \ion{Al}{III} $\lambda\lambda$1854,1862 doublets. Fig.~2 shows spectroscopic comparisons for each of the 10 BAL quasars listed in Table~3.\footnote{Fig.~2 displays four representative component figures from Fig.~2S, which can be viewed in the online, supplementary material (we hereafter only refer to Fig.~2). Fig.~2 consists of 14 component figures that display three comparisons from 122654--005430, two comparisons each from 140532+022947 and 221326+003846, and one comparison from the remaining sources (see Table~3).} We note that spectra for 025042+003536 and 104059+055524 presented in Fig.~2 do not allow for variability comparisons of the \ion{P}{V} BAL, however other spectra for these sources provide sufficient coverage to search for \ion{P}{V} BAL changes (see Table~1). Since there is no BAL variability information to present in these other cases, we therefore do not show all spectroscopic comparisons in Fig.~2. 

Previous studies focusing on \ion{C}{IV} and \ion{Si}{IV} BALs have concluded that BAL variability is common on rest frame time-scales of a few years and therefore a significant fraction of outflows exist within the vicinity of the SMBH (i.e. $\la$100\,pc, see e.g. \citealt{fil13}). \citet{fil13} measured \ion{C}{IV} and \ion{Si}{IV} BAL equivalent width variations in 291 BAL quasars, finding that approximately 50--60 per cent of the BALs vary over time-scales ranging from 1 to 3.7\,yr. \citet{cap13} concluded that over 50 per cent of their two-epoch comparisons exhibit \ion{C}{IV} BAL variability on 1-year time-scales, with a higher probability of detecting BAL changes over multi-year time-scales (see their fig.~9). Our measured BAL variability time-scales are broadly consistent with previous results. The low fraction of BAL quasars in our sample that exhibit significant \ion{C}{IV} BAL variations (i.e., $\sim$10 per cent) is consistent with previous studies which conclude that weaker \ion{C}{IV} BALs generally vary with greater frequency and with larger fractional EW changes than stronger \ion{C}{IV} BALs (see e.g., \citealt{cap11}; \citealt{fil13}); \ion{P}{V} BALs in our sample indicate a preference toward detecting strong \ion{C}{IV} BALs (see Fig.~2).

In some cases we observe variable BALs associated with more than a single ion increase or decrease together at similar velocities (see e.g., in Fig.~2, the \ion{C}{IV} and \ion{Si}{IV} BAL variations at \mbox{--9400}\,km\,s$^{-1}$ in 025042+003536 as well as BAL variations at roughly --5700\,km\,s$^{-1}$ from \ion{N}{iii}, \ion{S}{IV}, \ion{P}{V}, \ion{Si}{IV}, and \ion{C}{IV} in 114548+393746). We also detect BALs that change their variability direction from one comparison to the next. The BALs associated with \ion{N}{V}, \ion{Si}{IV}, and \ion{C}{IV} in 140532+022947 increase in strength over the first 2.4\,yr, then decrease in strength over the next 0.8\,yr (see Fig.~2). The \ion{C}{IV} BAL in 221326+003846 increases in strength over the first 2.0\,yr then decreases in strength over the next 0.8\,yr (see Fig.~2). The \ion{Si}{IV} BAL in 122654--005430  increases over the first 2.4\,yr and the next 0.8\,yr, and then decreases in strength over the next 0.5\,yr (see Fig.~2).

Understanding the origin of BAL variability is critical to constraining outflow energetics since the constraints we derive on the location of the absorber relative to the SMBH depend on the mechanism that produces the observed BAL variations. We utilize the spectroscopic comparisons in Fig.~2 in conjunction with CSS $V$-band light curves (representative cases are shown in Fig.~3) to interpret the BAL variations in our sample. Below we discuss variability scenarios that provide evidence supporting one of the two scenarios (see Section~1) as the plausible explanation for the detected BAL variability. The origin of BAL variations in most sources in our sample is ambiguous given the limited amount of information available and the unknown geometry of BAL outflows (see e.g., \citealt{ham04}), and we therefore only focus on sources with supporting evidence of one of the two scenarios in the subsequent discussion.

Detecting variability in a \ion{C}{IV} BAL that is highly optically thick supports models involving outflows crossing the LOS since a saturated absorber might not respond to fluctuations in the ionizing radiation field in a significant way (see Capellupo et al. 2014). The \ion{C}{IV} BAL is likely highly optically thick if \ion{P}{V} absorption exists at the same velocity (see Section 1), and we detect significant (Y) variations in 2 highly optically thick \ion{C}{IV} BALs with no detectable \ion{P}{V} BAL changes at the same velocity.\footnote{This behaviour can be explained with the transverse-motion scenario if the \ion{P}{V} BAL gas exhibits a different covering fraction and/or optical depth than the \ion{C}{IV} BAL gas, leading to \ion{P}{V} BAL changes that are not detectable given the available data.} We observe Y variations in the \ion{C}{IV} BAL in 221326+003846 over the first 2.0\,yr and the following 0.8\,yr, and \ion{P}{V} broad absorption exists at the same velocity as the observed \ion{C}{IV} BAL variability (see Fig.~2). Significant \ion{C}{IV} BAL variations are also detected with \ion{P}{V} broad absorption at similar velocities in 140532+022957 over a 0.8\,yr time-scale (see Fig.~2).

Observing variability in \ion{P}{V} and/or \ion{Si}{IV} BALs but not in highly optically thick \ion{C}{IV} broad absorption at the same velocity supports ionization-change scenarios, since transverse motions would likely alter a highly optically thick \ion{C}{IV} BAL profile (see above and Capellupo et al. 2014).\footnote{It is possible for such behaviour to occur under the transverse-motion scenario if different ions exhibit different covering fractions, and therefore could trace different parts of the outflowing gas (see discussion from \citealt{cap12}).} We detect Y variations in 6 \ion{P}{V} and/or \ion{Si}{IV} BALs with no detectable \ion{C}{IV} BAL changes at similar velocities. The quasar 013652+122501 exhibits a non-black, non-variable \ion{C}{IV} BAL that lies at velocities where P\thinspace\textsc{v} BAL variability is detected (see Fig.~2). Similarly, we observe a dramatic increase in the \ion{Si}{IV} BAL strength in 104059+055524 which exists at the same velocity as a non-black, non-variable \ion{C}{IV} BAL that is likely highly optically thick due to the presence of \ion{P}{V} broad absorption at the same velocity (see Fig.~2). The \ion{Si}{IV} BAL decreases in strength at --6800\,km\,s$^{-1}$ in 104247+061521, and this BAL lies at the same velocity as a non-variable \ion{C}{IV} BAL with associated \ion{P}{V} broad absorption at the same velocity (see Fig.~2). The \ion{Si}{IV} BAL decreases in strength at --2500\,km\,s$^{-1}$ over 0.5\,yr in 122654--005430, and this trough also lies at the same velocity as a non-variable \ion{C}{IV} BAL with \ion{P}{V} broad absorption at the same velocity (see Fig.~2).

We also see examples of significant, coordinated absorption line variations associated with \ion{C}{IV} over a large velocity interval that spans $>10\,000$\,km\,s$^{-1}$. Coordinated line variations over a large range of velocity seem to favour variations in the ionizing radiation field, however transverse-motion scenarios cannot be entirely ruled out (\citealt{ham11}; \citealt{cap13}; \citealt{gri15}; \citealt{wan15}). We observe decreases in \ion{C}{IV} absorption line depth from four distinct troughs over a \mbox{--14\,300}\,km\,s$^{-1}$ range between the DR7/DR12 epochs in 122654--005430, and from a single \ion{C}{IV} BAL over \mbox{-12\,600}\,km\,s$^{-1}$ between the DR7/DR12 observations for 140532+0222957 (see Fig.~2 and Table~3).

Fluctuations in the observed continuum flux provide evidence supporting an ionization change occurring within the absorber if a correlation exists between variations in continuum flux and ionizing radiation flux. We inspect the CSS light curves in combination with our detected BAL variability to examine whether any quasars exhibit significant $V$-band variations that are likely mostly due to continuum emission. Fig.~3 displays representative, synthesized $V$-band light curves for our variable (Y) BAL quasars. Three Y sources in our sample exhibit significant (i.e., $\gtrsim0.5$\,mag) CSS variations within the observation time that are likely mostly due to continuum emission (see 104247+061521, 122654--005430 and 221326+003845 in Fig.~3), however three Y objects show marginal (i.e., $\sim$0.2--0.5\,mag) CSS variations within the observation time (see 013652+122501, 114548+393746, and 140532+022957). The remaining four Y quasars exhibit poor sampling or small-amplitude (i.e., $\lesssim$0.2\,mag) variations in the CSS observations. In light of the mixed behaviour we observe in the CSS light curves across our Y sources, we do not find strong evidence for the sample as a whole supporting the ionization-change scenario.

\begin{figure*}
	\includegraphics[width=1.99\columnwidth]{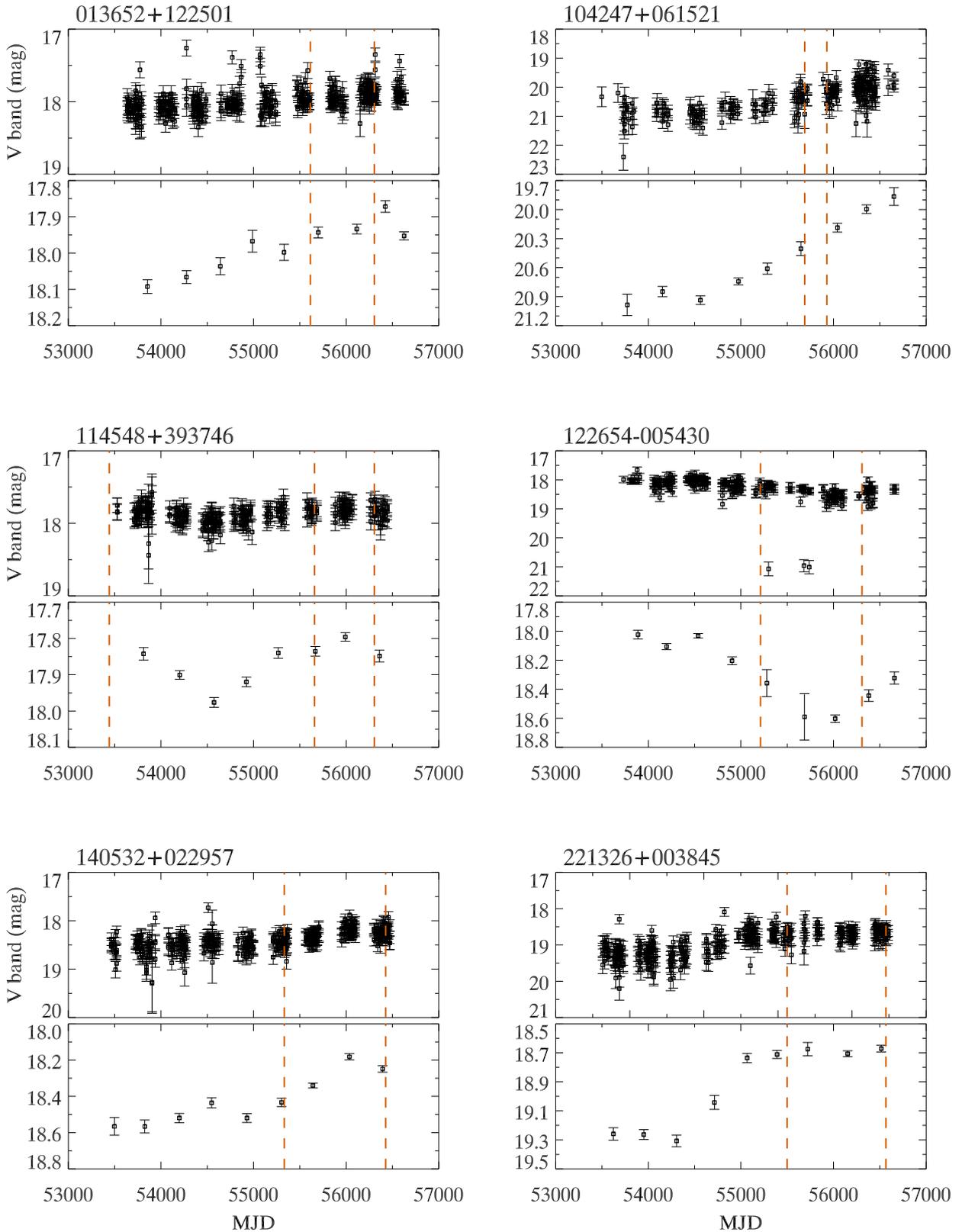}
	\caption{CSS $V$-band light curves, indicating flux variations. The source name is listed above each pair of panels. The upper panel for each source shows all photometric measurements and their errors, and each lower panel shows the mean $V$-band magnitude and standard deviation of the mean over one-year increments. Red dotted lines indicate the epochs of our spectroscopic observations (see Table~1).}
\end{figure*}

\subsection{Quasar outflow locations}

The BAL variations that we detect in 10 BAL quasars (see Fig.~2 and Table~3) collectively provide evidence supporting scenarios involving motions of gas across the LOS and ionization changes within the absorbers (see Section 4.1). Below we estimate representative BAL outflow distances from the SMBH using both interpretations and discuss our results in the context of previous studies that have constrained quasar outflow distances.

\subsubsection{Outflows crossing the LOS}

We consider outflowing gas crossing the LOS and use the Keplerian speed\footnote{\cite{mor17} argue that Keplerian crossing speeds behaving as $v_{\rmn{cross}} \propto 1/\sqrt{r}$ are a reasonable compromise for accretion disc winds where the flow geometry and acceleration mechanisms are poorly understood. In particular, the flows might expand freely with angular momentum conservation yielding a faster-than-Keplerian decline with radius $v_{\rmn{cross}} \propto 1/r$, or they might be influenced by magnetic fields threaded through the disc, as in magneto-centrifugal acceleration models \citep{eve02,fuk10} that lead to a slower decline in radius.} in order to obtain a characteristic upper limit for the outflow distance $r$ from the SMBH [i.e. \mbox{$r=GM_{\bullet}/v_{\rmn{cross}}^2$,} where $G$ is the gravitational constant, $M_{\bullet}$ is the SMBH mass (see Table 1), and $v_{\rmn{cross}}$ is the transverse speed of the absorber across the LOS].

We estimate the transverse speed $v_{\rmn{cross}}$ using the measured BAL variability time-scale $\Delta t$, the change in absorber covering fraction $\Delta A$, the diameter of the background light source at a specified wavelength $D_{\lambda}$, and the geometries of the absorber and background light source. We utilize an average BAL variability time-scale of $\Delta t \le1.22$\,yr for our calculation. The $\Delta A$ parameter is quantified as the average, normalized flux difference across the variable wavelength interval within the BAL (see \citealt{cap11}), and we assume $\tau>>1$ so the line depth is a direct measure of the covering fraction. We measure an average value of $\Delta A=0.18$ for our variable BALs, with an estimated uncertainty of 30 per cent [i.e., by using the 1$\sigma$ errors for each pair of spectra (see Section 3.1) and by assigning a 20 per cent error to the continuum flux values based on examination of results from alternative fits applied to the spectra].

The size of the background light source $D_{\lambda}$ is calculated using the standard thin-disc temperature profile in combination with Wien's law (see discussion in section 4.1 of \citealt{ede15}), and incorporates a modified constant to account for disc-size measurement discrepancies (Hamann et al. in preparation). The size $D_{\lambda}$ depends on $L_{\rmn{bol}}$ and $M_{\bullet}$ (see Table 1), the radiative efficiency $\eta$, and the wavelength $\lambda$. For our calculation we adopt log\,$L_{\rmn{bol}}=47.0$ and log\,$M_{\bullet}=9.4$ to be representative values for our sample, and the errors for $L_{\rmn{bol}}$ and $M_{\bullet}$ have been estimated to be $\sim$0.3 and 0.5 dex, respectively, based on systematic uncertainties in using single epoch spectra to measure black hole masses (see e.g., \citealt{she13}). Using the above measurements along with nominal values of $\eta=0.1$ and $\lambda=1320$\,\AA, we obtain a continuum source size estimate on the order of $D_{\lambda}=0.01$\,pc for our BAL quasar sample.

We adopt the knife-edge geometry from \citet{cap13} for our calculations of $v_{\rmn{cross}}$, which consists of a circular absorber of infinite radius crossing a square continuum source. The knife-edge model is used as a conservative, extreme geometry that should encompass the actual, unknown outflow geometries, and leads to a lower limit for the outflow crossing speed (i.e. $v_{\rmn{cross}} \ga \Delta A\,D_{\lambda}/\Delta t$) and therefore an upper limit on $r$. Our measurements yield a lower limit of $v_{\rmn{cross}}\ga1500$\,km\,s$^{-1}$ for typical variable BALs in our sample. Using the above considerations, we obtain an upper limit for the BAL outflow distance from the SMBH to be $r\sim$5~pc. Factoring in estimated uncertainties in the above parameters yields a range of distances on the order of $r\la1-10$\,pc for our sample using the Keplerian speed and transverse-motion interpretation. 

\subsubsection{Outflows undergoing an ionization change}

We consider an absorber undergoing an ionization change and constrain the outflow distance $r$ from the SMBH by considering an ionizing radiation field that illuminates a constant-density outflow [i.e. $r=\sqrt{L_{\rmn{ion}}/(4\pi U n_{\rmn{e}} c)}$, where $L_{\rmn{ion}}$ is the ionizing photon luminosity above 13.6\,eV, $U$ is the ionization parameter, $n_{\rmn{e}}$ is the electron density, and $c$ is the speed of light].

The ionizing photon luminosity $L_{\rmn{ion}}$ above 13.6\,eV is calculated by integrating a piece-wise power law ($L_{\nu}=\nu^{\alpha}$) for a typical quasar SED \citep{ham11} and using measurements of the luminosity at 1350\,\AA \ (see Section 2) to normalize the power law function. \citet{ham11} adopted  a quasar SED that consists of $\alpha=-1.0$ from 0.5 to 13.6\,eV, $\alpha=-2.56$ from 13.6 to 136\,eV, and $\alpha=-1.0$ from 136\,eV to 136\,keV. We obtain an average estimate, in photons\,s$^{-1}$, of log\,$L_{\rmn{ion}}=56.8$ for BAL quasars in our sample. 

\citet{wan15} analysed a large sample of quasars from SDSS DR10, concluding that coordinated BAL variations from \ion{C}{IV}, \ion{Si}{IV}, and \ion{N}{V} are statistically correlated with continuum variations in their sample. They attributed the variability in their sample to be broadly due to changes in the ionizing radiation field and further concluded that the majority of ions present in the absorbers are from higher ionization states than \ion{C}{IV}, \ion{Si}{IV}, and \ion{N}{V}. \citet{wan15} conducted photoionization models in conjunction with their results and constrained the ionization parameter for their BAL outflows to be log $U\ga0$. Based on the applicability of these results to BAL quasars more generally, we adopt the lower limit of log $U\ga0$ for our calculation.

The electron density is constrained by utilizing the measured BAL variability time-scale $\Delta t$ as an upper limit to the recombination time-scale of the gas; an upper limit on the recombination time yields a lower limit on the electron density (i.e. $n_{\rmn{e}} \ga 1/\alpha_{\rmn{rec}} \Delta t$, where $\alpha_{\rmn{rec}}$ represents the recombination rate coefficient). We calculate $\alpha_{\rmn{rec}}$ for the relevant ions of interest using version 8.0 of the CHIANTI atomic database (\citealt{der97}; \citealt{del15}), using a nominal temperature of 20 000 K for the BAL gas. We caution that constraining the electron density in this way does not account for the unknown ionization conditions in the BAL gas (see e.g., \citealt{ham97}), however our results are still useful for the purpose of obtaining order-of-magnitude limits on the outflow distance. We utilize average values of $\Delta t\le1.22$\,yr and $\alpha_{\rmn{rec}}=2.0\times10^{-11}$\,cm$^{3}$\,s$^{-1}$ to obtain a nominal lower limit of $n_{\rmn{e}}\ga1300$\,cm$^{-3}$. Using the above considerations and uncertainties, we estimate an upper limit for the BAL outflow distance from the SMBH to be $r\sim$350~pc and a range on the order of $r\la100-1000$\,pc using the ionization-change scenario.

\subsubsection{Comparisons to previous work}

Our BAL outflow distance constraints using the transverse-motion scenario (i.e., $r\la1-10$\,pc; see Section 4.2.1) and the ionization-change interpretation (i.e., $r\la100-1000$\,pc; see Section 4.2.2) indicate that we may be probing outflowing gas in close proximity to the SMBH, but the possibility remains that outflows in our sample exist at larger scales. We emphasize that these upper limits leaves open the possibility that the absorbers we probe are located at small distances from the SMBH under the ionization-change scenario. Below we compare our results to previous observational studies that have constrained quasar outflow distances.

Previous quasar outflow distance constraints using BAL variability studies are broadly consistent with our results. \citet{cap11,cap12,cap13} detected \ion{C}{IV} and \ion{Si}{IV} BAL variability and constrained the outflows in their sample to lie within $\sim$10\,pc of the SMBH by attributing the variations to absorbers crossing the LOS. \citet{cap13} also detected \ion{C}{IV} BAL changes on rapid time-scales, allowing them to constrain the absorber to lie between 0.001 and 0.02\,pc of the SMBH. Capellupo et al. (2014) found BAL variability associated with \ion{P}{V}, \ion{Si}{IV}, and \ion{C}{IV} at the same velocity in Q1413+1143, thereby providing strong support for absorbers crossing the LOS and allowed them to place an upper limit of $\sim$3.5\,pc for the outflow distance from the SMBH. Variability studies involving \ion{Fe}{II} and \ion{Mg}{II} BALs have constrained quasar outflows to lie within tens of parsecs of the SMBH by using the crossing absorber interpretation (\citealt{hal11}; \citealt{mcg15}). \citet{bar94} examined the variability patterns of \ion{C}{IV}, \ion{N}{V}, \ion{Si}{IV}, and \ion{Al}{III} BALs, yielding upper limits on the outflow distance to be a few hundred parsecs from the SMBH based on an ionization change scenario (see also \citealt{viv12}).

Studies of quasar outflows have also utilized photoionization models in conjunction with density-sensitive absorption lines to constrain outflow distances that are generally $\ga$100\,pc from the SMBH. Previous work using ground- and excited-state transitions in \ion{O}{IV} to constrain the gas density have estimated outflow distances of 3\,kpc \citep{ara13} from the SMBH. The \ion{S}{IV} ion also allows for density constraints, leading to outflow distances of $\sim$100\,pc (Chamberlain et al. 2015) and between 100 and 2000\,pc \citep{bor13} from the SMBH. Other studies have utilized the \ion{Si}{II} ion to estimate large outflow distances of 6\,kpc in an iron low-ionization BAL quasar \citep{dun10}. \citet{cha15b} utilized the \ion{N}{III} and \ion{S}{III} absorption lines in LBQS J1206+1052 to estimate the outflow distance to be $\sim$840\,pc from the SMBH.

In light of the above discussion, it is possible that quasar outflows exist at a range of distances from the SMBH. This conclusion suggests that quasar outflows might exhibit a range of energies and momenta and therefore some outflows may play a significant role in AGN feedback processes. We obtain nominal estimates on the energetics of BAL outflows in our sample (see Section 4.4) and discuss our results in the context of previous studies.   

\subsection{Quasar outflow hydrogen column densities}

We obtain an average lower limit on the \ion{P}{V} column density to be log\,$N_{\ion{P}{V}}\ga15.1$ based on our spectroscopic fits and integration routines (see Section 3.2), and consider this constraint to be representative of BAL outflows in our sample. \citet{ham98} carried out photoionization simulations of the BAL quasar PG 1254+047 and utilized his measured \ion{P}{V} column density lower limit of log $N_{\ion{P}{V}}=15.0$ to estimate the hydrogen column density of the flow to be log\,$N_{\rmn{H}}\ga22.0$, assuming solar abundances and constant-density clouds.

\citet{lei09} applied a template-fitting approach to the emerging BAL AGN WPVS 007 to measure a \ion{P}{V} column density of log $N_{\ion{P}{V}}\sim15.4$ and used photoionization models to estimate a hydrogen column density of log\,$N_{\rmn{H}}\ge22.2$ assuming solar abundances and an SED determined by using UV/optical spectra and $Swift$ X-ray observations. \citet{bor12} utilized a measured \ion{P}{V} column density of log\,$N_{\ion{P}{V}}\sim14.6$ in conjunction with \ion{S}{IV} to estimate a hydrogen column density of log\,$N_{\rmn{H}}=21.9$ in the BAL quasar SDSS J1512+1119. 

In light of previous results, we adopt the ratio log\,$N_{\rmn{H}}$/log\,$N_{\ion{P}{V}}>7.0$ from \citet{ham98} to estimate a nominal, average lower limit of log\,$N_{\rmn{H}}\ga22.1$ for BAL outflows in our sample. Of the previous studies above that place observational constraints on outflow column densities, we adopt the \citet{ham98} parameters to use as representative values for obtaining order-of-magnitude estimates of outflow energetics. The nominal values we adopt are consistent with Leighly et al. (2011) who analysed a range of BAL outflow ionization conditions using photoionization models, and concluded that the \ion{P}{V} lines do not become prominent until the absorber exhibits high column densities on the order of log $N_{\rmn{H}}\sim$22.0--23.0 assuming solar abundances (see their fig.~15).

\subsection{Quasar outflow energetics}

We quantify BAL outflow energetics by calculating the mass flow rate $\dot{M}_{\rmn{out}}$, kinetic luminosity $L_{\rmn{k}}$, and momentum flux $\dot{P}$ for variable (Y) BALs in our sample. These parameters depend on measurements of the outflow LOS velocity $v_{\textrm{LOS}}$ (see Table 3), outflow hydrogen column density $N_{\rmn{H}}$, outflow distance $r$ from the SMBH (see Sections 4.2.1 and 4.2.2), and outflow global covering fraction $Q$. For the subsequent calculations we adopt a value of log\,$N_{\rmn{H}}=22.1$ based on our \ion{P}{V} column density measurements (see Sections 3.2 and 4.3), a nominal value of $Q=15$ per cent based on the observed fraction of BAL quasars (see e.g., \citealt{tru06}), and a representative value of $v_{\rm{LOS}}=10\,000\,$\,km\,s$^{-1}$ based on our BAL velocity measurements.

The average mass flow rate is calculated using a characteristic flow time (i.e., $t_{\rmn{flow}}=r/v_{\rmn{LOS}}$) along with an estimate of the outflow mass $M_{\rmn{out}}$, and is quantified as \mbox{$\dot{M}_{\rmn{out}}=M_{\rmn{out}}/t_{\rmn{flow}}$.} The mass of the outflow is obtained  by assuming a thin-shell geometry and is calculated using the following equation: \\ 
\begin{equation}
    M_{\rmn{out}} \approx 4100 \Big( \frac{Q}{15\%} \Big) \Big(\frac{N_{\rmn{H}}}{2\times10^{22}\,\rmn{cm}^{-2}} \Big) \Big(\frac{r}{3.5 \rmn{pc}} \Big)^2 \ M_{\odot}
\end{equation} \\
where the parameters are defined above [see equation (2) from Capellupo et al. 2014]. For comparison we also estimate the accretion rate {$\dot{M}_{\rmn{acc}}$ using our representative value of $L_{\rmn{bol}}=47.0$ and a radiative efficiency of $\eta=$0.1 [i.e. \mbox{$\dot{M}_{\rmn{acc}}=L_{\rmn{bol}}/(\eta c^2$)].} Using the above parameters we calculate the kinetic luminosity [$L_{\rmn{k}}=(1/2)\dot{M}_{\rmn{out}} v_{\textrm{LOS}}^2$] and momentum flux ($\dot{P}=\dot{M}_{\rmn{out}} v_{\textrm{LOS}}$) in units of $L_{\rmn{bol}}$ and $L_{\rmn{bol}}/c$, respectively. We utilize the above values and consider a BAL outflow located at $r=1,10,100,1000$\,pc from the SMBH, consistent with representative distances we estimate using the transverse-motion and ionization-change interpretations, and calculate outflow energetics using these four nominal locations; Table~4 lists the results.

\begin{table*}
	\centering
	\caption{BAL outflow energetics}
	\begin{tabular}{cccccccc}
		\hline
		$r$ & $M_{\rmn{out}}$ & $t_{\rmn{flow}}$ & $\dot{M}_{\rmn{out}}$ & $\dot{M}_{\rmn{acc}}$ & $L_{\rmn{k}}$ & $\dot{P}$ \\
		(pc) & ($M_{\odot}$) & (yr) & ($M_{\odot}\,\rmn{yr}^{-1}$) & ($M_{\odot}\,\rmn{yr}^{-1}$) & ($L_{\rmn{bol}}$) & ($L_{\rmn{bol}}/c$)\\
		\hline
		1 & 100 & 100 & 1 & 10 & 0.001 & 0.01 \\
		10 & 1$\times10^{4}$ & 1000 & 10 & 10 & 0.01 & 0.1 \\ 
		100 & 1$\times10^{6}$ & 1$\times10^{4}$ & 100 & 10 & 0.1 & 1 \\
		1000 & 1$\times10^{8}$ & 1$\times10^{5}$ & 1000 & 10 & 1 & 10 \\
		\hline
	\end{tabular}\\
	Information listed for each quasar includes (from left to right): Assumed BAL outflow distance from the SMBH, outflow mass, characteristic flow time, mass flow rate, mass accretion rate, kinetic luminosity (in units of $L_{\rmn{bol}}$), and momentum flux (in units of $L_{\rmn{bol}}/c$). 
\end{table*}

In order to assess the viability of the outflows in our sample to contribute to AGN feedback it is necessary to make comparisons to proposed thresholds for outflows to become energetic enough to affect the host galaxy. Previous work has generally concluded that an outflow with a kinetic luminosity of $\sim$0.05 $L_{\rmn{bol}}$ can significantly accelerate ISM gas (see e.g., \citealt{kin15}). \citet{hop10} noted that the above threshold assumes that the outflowing gas affects all phases of the ISM, leading them to propose decreases in the above threshold to $\sim$0.005 $L_{\rmn{bol}}$ by arguing that outflows need only drive the hot, diffuse gas to affect host galaxy properties.

Our estimates of $L_{\rmn{k}}$ listed in Table~4 are crude approximations for the outflow energetics due to their large uncertainties, partly because the calculations involve multiplying hydrogen column density lower limits by outflow distance upper limits. However our order-of-magnitude energetics estimates using a range of distances can be broadly compared to the proposed $\sim$0.005--0.05 $L_{\rmn{bol}}$ thresholds outlined above. If a significant number of the BAL outflows in our sample exists in close proximity to the SMBH (i.e., within 10\,pc), then it is possible that these winds may only marginally contribute to AGN feedback processes. Conversely if a large number of BAL outflows are located much farther from the SMBH (i.e., 100--1000\,pc), then it is possible that these winds would contain sufficient amounts of energy and momentum to contribute to feedback.

Our nominal kinetic luminosity estimates in Table~4 range from 0.001 to 1 $L_{\rmn{bol}}$, which is broadly consistent with previous studies that utilized independent methodologies to estimate outflow energetics. Other studies have estimated kinetic luminosities on the order of  0.001 \citep{dun10} and 0.01 $L_{\rmn{bol}}$ (\citealt{moe09}; \citealt{ara13}; Capellupo et al. 2014). Larger values of $L_{\rmn{k}}$ have been estimated near the proposed threshold for feedback of 0.05 $L_{\rmn{bol}}$ \citep{bor13}. Our results indicate that many BAL outflows in our sample possess kinetic energies that are viable for AGN feedback processes.

\section{Conclusions}

We analyse multiple-epoch spectra of 71 BAL quasars that exhibit conservative detections of the \ion{P}{V} $\lambda\lambda$1117,1128 BAL, and find significant variability in 10 sources using quantitative criteria. Our measured rest frame BAL variability time-scales range from $\le0.2-3.8$\,yr, and variable wavelength intervals are associated with a range of ions including \ion{N}{III}, \ion{O}{VI}, \ion{S}{IV}, \ion{P}{V}, \ion{H}{I}, \ion{N}{V}, \ion{Si}{IV}, and \ion{C}{IV}. Our results are consistent with previous work with regard to BAL variability time-scale, velocity, and direction.

We interpret the variable BALs in our sample and find sources that provide evidence supporting transverse motions of gas or ionization changes as explanations for the origin of the detected BAL changes (see Section 4.1). In particular, detecting significant variability in a saturated \ion{C}{IV} BAL is evidence supporting an absorber crossing the LOS. We also find evidence supporting an ionization change scenario by detecting significant variability in only the \ion{P}{V} and/or \ion{Si}{IV} BAL and by identifying coordinated \ion{C}{IV} absorption line variability over large velocity intervals. Future high-resolution observations focusing on absorption lines from multiple ions and doublets with large wavelength separations such as the \ion{P}{V} $\lambda\lambda$1117,1128, \ion{Si}{IV} $\lambda\lambda$1393,1402, and \ion{Al}{III} $\lambda\lambda$1854,1862 BALs will be useful in probing the origin of BAL changes over localized velocity intervals.

Using simple models involving an outflow crossing the LOS, we conclude that BAL outflows in our sample nominally exist on the order of $r\la1-10$\,pc from the SMBH. Under the ionization change interpretation, the BAL outflows we probe are constrained to exist on the order of $r\la100-1000$\,pc from the SMBH. Obtaining multiple-epoch spectra over shorter rest frame time-scales will be necessary for placing more reliable constraints on outflow locations, which will have important implications for the derived feedback-related quantities.

 We utilize our quasar outflow distance constraints in conjunction with hydrogen column density limits to estimate kinetic luminosities ranging from 0.001 to 1 times $L_{\rmn{bol}}$ for outflows in our sample, implying that many quasar outflow energies in our sample are viable for AGN feedback. Our results provide a significant increase in the available information on quasar outflow energetics. 
 
 Future studies that focus on absorbers with a range of distances and column densities will be important to better understand the role of quasar outflows in AGN feedback. This work probes high column density outflows using the \ion{P}{V} BAL and constrains outflowing gas locations using the BAL variability technique. Utilizing the \ion{S}{IV} and \ion{Al}{III} BALs in particular will trace a broader range of column densities, and will give us a more complete picture of quasar outflow demographics when used in conjunction with BAL variability studies.

\section*{Acknowledgements}

This work is based on observations obtained at the MDM Observatory, operated by Dartmouth College, Columbia University, Ohio State University, Ohio University, and the University of Michigan.

Funding for SDSS-III has been provided by the Alfred P. Sloan Foundation, the Participating Institutions, the National Science Foundation, and the U.S. Department of Energy Office of Science. The SDSS-III web site is http://www.sdss3.org/.

SDSS-III is managed by the Astrophysical Research Consortium for the Participating Institutions of the SDSS-III Collaboration including the University of Arizona, the Brazilian Participation Group, Brookhaven National Laboratory, Carnegie Mellon University, University of Florida, the French Participation Group, the German Participation Group, Harvard University, the Instituto de Astrofisica de Canarias, the Michigan State/Notre Dame/JINA Participation Group, Johns Hopkins University, Lawrence Berkeley National Laboratory, Max Planck Institute for Astrophysics, Max Planck Institute for Extraterrestrial Physics, New Mexico State University, New York University, Ohio State University, Pennsylvania State University, University of Portsmouth, Princeton University, the Spanish Participation Group, University of Tokyo, University of Utah, Vanderbilt University, University of Virginia, University of Washington, and Yale University.

The CSS survey is funded by the National Aeronautics and Space Administration under Grant No. NNG05GF22G issued through the Science Mission Directorate Near-Earth Objects Observations Program.  The CRTS survey is supported by the U.S.~National Science Foundation under grants AST-0909182 and AST-1313422.

CHIANTI is a collaborative project involving George Mason University, the University of Michigan (USA) and the University of Cambridge (UK).

This research has made use of data obtained from the Chandra Source Catalog, provided by the Chandra X-ray Center (CXC) as part of the Chandra Data Archive.

\appendix

\section{Notes on Individual Quasars} 

Below we briefly discuss 3 out of 10 BAL quasars that exhibit significant (Y) BAL variability in our sample (see Section 3.1 and Table~3). These three sources exhibit absorption-line changes that are the least convincing of our Y sources, and we outline reasons why the detected variations are plausibly real; the Y quasars not mentioned here show convincing absorption-line variability. \\

\emph{013652+122501} \medskip

We detect significant variability in the \ion{P}{V} BAL at --6200\,km\,s\,$^{-1}$ between the DR12 and MDM observations (see Fig. 2).\footnote{Note that, when referring to Fig. 2, the velocity shift in the text corresponds to the velocity of the identified ionic transitions in the associated plot. Identified velocities are chosen to call attention to important regions that exhibit BAL variability.} The variability is localized (i.e. occurs over a $\sim$10\,\AA \ wide region) and is therefore likely not due to fluctuations in continuum emission that would exhibit changes over a large portion of the wavelength coverage (i.e. $\gtrsim$1000\,\AA). The variability is located at wavelengths that are probably not associated with nearby BELs that could also produce significant variations. The variable wavelength interval is therefore reasonably due to absorption. The flux change is plausibly real since it is greater than 4$\sigma$ over several pixels, is not contaminated by night sky lines, and we observe good agreement between the spectra on either side of the \ion{P}{V} BAL (i.e., between --11\,000 and --9000\,km\,s$^{-1}$ and between --2000 and 0\,km\,s$^{-1}$). \\

\emph{102154+051646}

We detect a significant \ion{C}{IV} absorption line depth increase at --23\,900\,km\,s$^{-1}$ between the DR7 and DR12 epochs (see Fig.~2). This flux change is localized and lies well outside regions that would be confused by BEL changes, and exists within a high-velocity \ion{C}{IV} trough. The flux variations is above 4$\sigma$ over many consecutive pixels and we observe good agreement on either side of the variable interval (i.e., less than --25\,000\,km\,s$^{-1}$ and greater than --21\,000\,km\,s$^{-1}$). 

The \ion{C}{IV} BAL centered at approximately \mbox{--6000\,km\,s$^{-1}$} appears saturated, and there is likely \ion{P}{V} broad absorption at similar velocities (see Fig.~2). It is unclear whether the observed \ion{C}{IV} absorption line increase at --23\,900\,km\,s$^{-1}$ has associated \ion{P}{V} absorption; the broad absorption feature centered at --20\,000\,km\,s$^{-1}$ in the \ion{P}{V} velocity plot in Fig.~2 is likely attributed to a \ion{S}{IV} $\lambda\lambda$1062,1073 BAL. \\

\emph{221326+003846}

The \ion{C}{IV} BAL exhibits a significant flux increase at \mbox{--10\,000}\,km\,s$^{-1}$ between the DR7 and DR12 observations, followed by a flux decrease at --8100\,km\,s$^{-1}$ between the DR12 and MDM epochs (see Fig.~2). Both of these variable regions are localized and within a \ion{C}{IV} BAL that is significantly detached from the nearby \ion{C}{IV} BEL, indicating that the variations are likely due to absorption. Both variations are apparent by eye (i.e., due to the the good agreement in flux on either side of the \ion{C}{IV} BAL), and exhibit flux differences that are greater than 4$\sigma$ over several consecutive pixels.

\section{X-ray observations of 122654--005430}

X-ray spectroscopic observations are useful for studying the highly-ionized component of quasar outflows that are not traced by rest frame UV BALs. One source in our sample, 122654--005430, was observed on MJD 52708 using the ACIS instrument onboard the $Chandra$ X-ray Observatory, and we use these observations to extract information on the X-ray spectroscopic shape and degree of absorption in this source. 122654--005430 was observed for 20\,800\,s and 49 source counts were detected over an energy range of \mbox{0.5--7}\,keV.

The X-ray spectroscopic shape and strength of absorption in 122654--005430 is characterized by comparing observed hardness ratios (H--M)/(H+M) and (M--S)/(M+S)\footnote{The H, M, and S energy bands range from 2.0--7.0, 1.2--2.0, and 0.5--1.2\,keV, respectively.} to predicted values to determine the power law photon index $\Gamma$ (i.e. $f_{\rmn{E}}\propto E^{-\Gamma}$) and intrinsic absorber column density $N_{\rmn{H}}$. Observed hardness ratios were calculated using the bayesian code from \citet{par06} and predicted hardness ratios were generated using the $Chandra$ Portable, Interactive Multi-Mission Simulator (PIMMS) tool. Predicted hardness ratios were determined by varying $\Gamma$ and $N_{\rmn{H}}$ and were normalized using the H band count rate for \mbox{122654--005430}. Our predictions include a foreground absorber with column density 2.1$\times10^{20}$\,cm$^{-2}$ that was calculated using the Galactic \ion{H}{I} maps from \citet{dic90}.

From the observations we calculate hardness ratios for 122654--005430 to be (H--M)/(H+M)$=-0.51^{+0.15}_{-0.16}$ and \mbox{(M--S)/(M+S)}$=0.19^{+0.16}_{-0.16}$ at 95 per cent confidence, which is consistent with a photon index of $\Gamma\sim3$ and an intrinsic absorber column density of log $N_{\rmn{H}}\sim23$. Our value of $\Gamma\sim3$ is steeper than results from \citet{gal02}, who measured an average X-ray continuum index of $\Gamma\sim2$ for their BAL quasar sample. \citet{geo00} obtained a range of X-ray slopes between $1.5\la \Gamma \la3$ and an average index of $\Gamma\sim2$ for their sample of non-BAL quasars. Our extracted intrinsic column density of log $N_{\rmn{H}}\sim23$ is consistent with previous work that derived X-ray absorber column densities between log $N_{\rmn{H}}\sim22-24$ for BAL quasars \citep{gal06}.

We also investigate the relationship between X-ray and UV flux by calculating the difference between the observed and predicted values of the two point optical-to-X-ray power law index $\alpha_{\rmn{ox}}$ [i.e. $\Delta\alpha_{\rmn{ox}}=\alpha_{\rmn{ox}}(\rmn{observed})-\alpha_{\rmn{ox}}(\rmn{predicted})$]. The observed value of $\alpha_{\rmn{ox}}$ is defined as 0.384 log ($f_{\rmn{2\,keV}}/f_{\rmn{2500\,A}})$, where $f_{\rm{2\,keV}}$ and $f_{\rmn{2500\,A}}$ are the flux densities at rest-frame 2\,keV and 2500\,\AA, respectively. The flux density at 2\,keV is a measure of the absorption-free continuum and was determined with PIMMS using the H band count rate to estimate the flux density at 7\,keV and assuming a power law index $\Gamma=3$. The predicted value of $\alpha_{\rmn{ox}}$ is estimated from the $\alpha_{\rmn{ox}}-L_{\rmn{2500\,A}}$ relation for non-BAL quasars from \citet{str05} [see their equation (6)], where $L_{\rmn{2500\,A}}$ is the luminosity density at 2500\,\AA. For the purpose of making comparisons to \citet{gal06}, we do not correct our measured $f_{2500\,A}$ value for Galactic extinction; reddening due to foreground absorbers has the effect of decreasing $\alpha_{\rmn{ox}}$ and $\Delta\alpha_{\rmn{ox}}$ by up to $\sim$0.09 and 0.06, respectively (see discussion in section 3 of \citealt{gal06}).

We calculate $\alpha_{\rmn{ox}}$ and $\Delta\alpha_{\rmn{ox}}$ values of --1.35 and +0.33, respectively, for 122654--005430. Our result is larger than \citet{gal06} who derived an average value of $\Delta\alpha_{\rmn{ox}}(\rmn{corr})\sim-0.10$ for their sample of 35 BAL quasars. \citet{gib09} quantified $\Delta\alpha_{\rmn{ox}}$ for 166 non-BAL quasars to be on average 0.00$\pm$0.01, which includes corrections for Galactic extinction at rest-frame 2500\,\AA. We conclude that 122654--005430 exhibits X-ray absorption consistent with previous studies of BAL quasars, but shows a ratio of X-ray-to-UV flux levels that is larger than typically found in previous studies of BAL and non-BAL quasars.

% Don't change these lines
\bsp	% typesetting comment
\label{lastpage}
\end{document}